\documentclass[twocolumn,letterpaper,aps,prc,superscriptaddress,showpacs,floatfix]{revtex4}

\usepackage{graphicx}   

\begin{document}

\title{Spectra and ratios of identified particles in Au$+$Au and $d$$+$Au collisions 
at $\sqrt{s_{_{NN}}}$~=~200~GeV}

\newcommand{\abilene}{Abilene Christian University, Abilene, Texas 79699, USA}
\newcommand{\augie}{Department of Physics, Augustana College, Sioux Falls, South Dakota 57197, USA}
\newcommand{\banaras}{Department of Physics, Banaras Hindu University, Varanasi 221005, India}
\newcommand{\barc}{Bhabha Atomic Research Centre, Bombay 400 085, India}
\newcommand{\baruch}{Baruch College, City University of New York, New York, New York, 10010 USA}
\newcommand{\bnlcoll}{Collider-Accelerator Department, Brookhaven National Laboratory, Upton, New York 11973-5000, USA}
\newcommand{\bnlphys}{Physics Department, Brookhaven National Laboratory, Upton, New York 11973-5000, USA}
\newcommand{\caucr}{University of California - Riverside, Riverside, California 92521, USA}
\newcommand{\charlesczech}{Charles University, Ovocn\'{y} trh 5, Praha 1, 116 36, Prague, Czech Republic}
\newcommand{\chonbuk}{Chonbuk National University, Jeonju, 561-756, Korea}
\newcommand{\ciae}{Science and Technology on Nuclear Data Laboratory, China Institute of Atomic Energy, Beijing 102413, P.~R.~China}
\newcommand{\cns}{Center for Nuclear Study, Graduate School of Science, University of Tokyo, 7-3-1 Hongo, Bunkyo, Tokyo 113-0033, Japan}
\newcommand{\colorado}{University of Colorado, Boulder, Colorado 80309, USA}
\newcommand{\columbia}{Columbia University, New York, New York 10027 and Nevis Laboratories, Irvington, New York 10533, USA}
\newcommand{\czechtech}{Czech Technical University, Zikova 4, 166 36 Prague 6, Czech Republic}
\newcommand{\dapnia}{Dapnia, CEA Saclay, F-91191, Gif-sur-Yvette, France}
\newcommand{\debrecen}{Debrecen University, H-4010 Debrecen, Egyetem t{\'e}r 1, Hungary}
\newcommand{\elte}{ELTE, E{\"o}tv{\"o}s Lor{\'a}nd University, H - 1117 Budapest, P{\'a}zm{\'a}ny P. s. 1/A, Hungary}
\newcommand{\ewha}{Ewha Womans University, Seoul 120-750, Korea}
\newcommand{\fit}{Florida Institute of Technology, Melbourne, Florida 32901, USA}
\newcommand{\fsu}{Florida State University, Tallahassee, Florida 32306, USA}
\newcommand{\gsu}{Georgia State University, Atlanta, Georgia 30303, USA}
\newcommand{\hiroshima}{Hiroshima University, Kagamiyama, Higashi-Hiroshima 739-8526, Japan}
\newcommand{\ihepprot}{IHEP Protvino, State Research Center of Russian Federation, Institute for High Energy Physics, Protvino, 142281, Russia}
\newcommand{\illuiuc}{University of Illinois at Urbana-Champaign, Urbana, Illinois 61801, USA}
\newcommand{\inrras}{Institute for Nuclear Research of the Russian Academy of Sciences, prospekt 60-letiya Oktyabrya 7a, Moscow 117312, Russia}
\newcommand{\instpasczech}{Institute of Physics, Academy of Sciences of the Czech Republic, Na Slovance 2, 182 21 Prague 8, Czech Republic}
\newcommand{\isu}{Iowa State University, Ames, Iowa 50011, USA}
\newcommand{\jaea}{Advanced Science Research Center, Japan Atomic Energy Agency, 2-4 Shirakata Shirane, Tokai-mura, Naka-gun, Ibaraki-ken 319-1195, Japan}
\newcommand{\jinrdubna}{Joint Institute for Nuclear Research, 141980 Dubna, Moscow Region, Russia}
\newcommand{\jyvaskyla}{Helsinki Institute of Physics and University of Jyv{\"a}skyl{\"a}, P.O.Box 35, FI-40014 Jyv{\"a}skyl{\"a}, Finland}
\newcommand{\kek}{KEK, High Energy Accelerator Research Organization, Tsukuba, Ibaraki 305-0801, Japan}
\newcommand{\korea}{Korea University, Seoul, 136-701, Korea}
\newcommand{\kurchatov}{Russian Research Center ``Kurchatov Institute", Moscow, 123098 Russia}
\newcommand{\kyoto}{Kyoto University, Kyoto 606-8502, Japan}
\newcommand{\labllr}{Laboratoire Leprince-Ringuet, Ecole Polytechnique, CNRS-IN2P3, Route de Saclay, F-91128, Palaiseau, France}
\newcommand{\lahorelums}{Physics Department, Lahore University of Management Sciences, Lahore, Pakistan}
\newcommand{\lawllnl}{Lawrence Livermore National Laboratory, Livermore, California 94550, USA}
\newcommand{\losalamos}{Los Alamos National Laboratory, Los Alamos, New Mexico 87545, USA}
\newcommand{\lpc}{LPC, Universit{\'e} Blaise Pascal, CNRS-IN2P3, Clermont-Fd, 63177 Aubiere Cedex, France}
\newcommand{\lund}{Department of Physics, Lund University, Box 118, SE-221 00 Lund, Sweden}
\newcommand{\maryland}{University of Maryland, College Park, Maryland 20742, USA}
\newcommand{\mass}{Department of Physics, University of Massachusetts, Amherst, Massachusetts 01003-9337, USA }
\newcommand{\michigan}{Department of Physics, University of Michigan, Ann Arbor, Michigan 48109-1040, USA}
\newcommand{\muenster}{Institut fur Kernphysik, University of Muenster, D-48149 Muenster, Germany}
\newcommand{\muhlenberg}{Muhlenberg College, Allentown, Pennsylvania 18104-5586, USA}
\newcommand{\myongji}{Myongji University, Yongin, Kyonggido 449-728, Korea}
\newcommand{\nagasaki}{Nagasaki Institute of Applied Science, Nagasaki-shi, Nagasaki 851-0193, Japan}
\newcommand{\newmex}{University of New Mexico, Albuquerque, New Mexico 87131, USA }
\newcommand{\nmsu}{New Mexico State University, Las Cruces, New Mexico 88003, USA}
\newcommand{\ohio}{Department of Physics and Astronomy, Ohio University, Athens, Ohio 45701, USA}
\newcommand{\ornl}{Oak Ridge National Laboratory, Oak Ridge, Tennessee 37831, USA}
\newcommand{\orsay}{IPN-Orsay, Universite Paris Sud, CNRS-IN2P3, BP1, F-91406, Orsay, France}
\newcommand{\peking}{Peking University, Beijing 100871, P.~R.~China}
\newcommand{\pnpi}{PNPI, Petersburg Nuclear Physics Institute, Gatchina, Leningrad region, 188300, Russia}
\newcommand{\riken}{RIKEN Nishina Center for Accelerator-Based Science, Wako, Saitama 351-0198, Japan}
\newcommand{\rikjrbrc}{RIKEN BNL Research Center, Brookhaven National Laboratory, Upton, New York 11973-5000, USA}
\newcommand{\rikkyo}{Physics Department, Rikkyo University, 3-34-1 Nishi-Ikebukuro, Toshima, Tokyo 171-8501, Japan}
\newcommand{\saispbstu}{Saint Petersburg State Polytechnic University, St. Petersburg, 195251 Russia}
\newcommand{\saopaulo}{Universidade de S{\~a}o Paulo, Instituto de F\'{\i}sica, Caixa Postal 66318, S{\~a}o Paulo CEP05315-970, Brazil}
\newcommand{\seoulnat}{Seoul National University, Seoul, Korea}
\newcommand{\stonybrkc}{Chemistry Department, Stony Brook University, SUNY, Stony Brook, New York 11794-3400, USA}
\newcommand{\stonycrkp}{Department of Physics and Astronomy, Stony Brook University, SUNY, Stony Brook, New York 11794-3400, USA}
\newcommand{\tenn}{University of Tennessee, Knoxville, Tennessee 37996, USA}
\newcommand{\titech}{Department of Physics, Tokyo Institute of Technology, Oh-okayama, Meguro, Tokyo 152-8551, Japan}
\newcommand{\tsukuba}{Institute of Physics, University of Tsukuba, Tsukuba, Ibaraki 305, Japan}
\newcommand{\vandy}{Vanderbilt University, Nashville, Tennessee 37235, USA}
\newcommand{\waseda}{Waseda University, Advanced Research Institute for Science and Engineering, 17 Kikui-cho, Shinjuku-ku, Tokyo 162-0044, Japan}
\newcommand{\weizmann}{Weizmann Institute, Rehovot 76100, Israel}
\newcommand{\wigner}{Institute for Particle and Nuclear Physics, Wigner Research Centre for Physics, Hungarian Academy of Sciences (Wigner RCP, RMKI) H-1525 Budapest 114, POBox 49, Budapest, Hungary}
\newcommand{\yonsei}{Yonsei University, IPAP, Seoul 120-749, Korea}
\affiliation{\abilene}
\affiliation{\augie}
\affiliation{\banaras}
\affiliation{\barc}
\affiliation{\baruch}
\affiliation{\bnlcoll}
\affiliation{\bnlphys}
\affiliation{\caucr}
\affiliation{\charlesczech}
\affiliation{\chonbuk}
\affiliation{\ciae}
\affiliation{\cns}
\affiliation{\colorado}
\affiliation{\columbia}
\affiliation{\czechtech}
\affiliation{\dapnia}
\affiliation{\debrecen}
\affiliation{\elte}
\affiliation{\ewha}
\affiliation{\fit}
\affiliation{\fsu}
\affiliation{\gsu}
\affiliation{\hiroshima}
\affiliation{\ihepprot}
\affiliation{\illuiuc}
\affiliation{\inrras}
\affiliation{\instpasczech}
\affiliation{\isu}
\affiliation{\jaea}
\affiliation{\jinrdubna}
\affiliation{\jyvaskyla}
\affiliation{\kek}
\affiliation{\korea}
\affiliation{\kurchatov}
\affiliation{\kyoto}
\affiliation{\labllr}
\affiliation{\lahorelums}
\affiliation{\lawllnl}
\affiliation{\losalamos}
\affiliation{\lpc}
\affiliation{\lund}
\affiliation{\maryland}
\affiliation{\mass}
\affiliation{\michigan}
\affiliation{\muenster}
\affiliation{\muhlenberg}
\affiliation{\myongji}
\affiliation{\nagasaki}
\affiliation{\newmex}
\affiliation{\nmsu}
\affiliation{\ohio}
\affiliation{\ornl}
\affiliation{\orsay}
\affiliation{\peking}
\affiliation{\pnpi}
\affiliation{\riken}
\affiliation{\rikjrbrc}
\affiliation{\rikkyo}
\affiliation{\saispbstu}
\affiliation{\saopaulo}
\affiliation{\seoulnat}
\affiliation{\stonybrkc}
\affiliation{\stonycrkp}
\affiliation{\tenn}
\affiliation{\titech}
\affiliation{\tsukuba}
\affiliation{\vandy}
\affiliation{\waseda}
\affiliation{\weizmann}
\affiliation{\wigner}
\affiliation{\yonsei}
\author{A.~Adare} \affiliation{\colorado}
\author{S.~Afanasiev} \affiliation{\jinrdubna}
\author{C.~Aidala} \affiliation{\mass} \affiliation{\michigan}
\author{N.N.~Ajitanand} \affiliation{\stonybrkc}
\author{Y.~Akiba} \affiliation{\riken} \affiliation{\rikjrbrc}
\author{H.~Al-Bataineh} \affiliation{\nmsu}
\author{J.~Alexander} \affiliation{\stonybrkc}
\author{A.~Angerami} \affiliation{\columbia}
\author{K.~Aoki} \affiliation{\kyoto} \affiliation{\riken}
\author{N.~Apadula} \affiliation{\stonycrkp}
% ...
\collaboration{PHENIX Collaboration} \noaffiliation

\date{\today}

\begin{abstract}
The transverse momentum ($p_T$) spectra and ratios of identified charged hadrons
($\pi^{\pm}$, $K^{\pm}$, $p$, $\bar{p}$)
produced in $\sqrt{s_{_{NN}}}$~=~200~GeV Au+Au and d+Au collisions are reported in
five different centrality classes for each collision species.  The measurements of
pions and protons are reported up to $p_T$~=~6~GeV/c (5~GeV/c), and the
measurements of kaons are reported up to $p_T$~=~4~GeV/c (3.5~GeV/c) in Au$+$Au
($d$$+$Au) collisions.  In the intermediate $p_T$ region, between 2--5~GeV/c, a
significant enhancement of baryon to meson ratios compared to those measured in
$p$$+$$p$ collisions is observed.  This enhancement is present in both Au+Au and
d+Au collisions, and increases as the collisions become more central.  We
compare a class of peripheral Au+Au collisions with a class central d+Au
collisions which have a comparable number of participating nucleons and binary
nucleon-nucleon collisions.  The $p_T$ dependent particle ratios for these
classes display a remarkable similarity, which is then discussed.

\end{abstract}


\maketitle


\section{Introduction}

Measurements of identified particles in Au$+$Au collisions allow
the study of particle production mechanisms in a hot and dense
nuclear medium and probe the properties of the quark-gluon plasma
(QGP)~\cite{Adcox:2004mh,Adams:2005dq,Back:2004je,Arsene:2004fa}.
In $d$$+$Au collisions, these measurements allow the study of
cold nuclear matter effects on particle production,
such as the Cronin enhancement~\cite{Cronin:1974zm,Antreasyan:1978cw},
nuclear shadowing~\cite{Arneodo:1992wf},
and gluon saturation~\cite{Iancu:2003xm}.
These cold nuclear matter effects are present in Au$+$Au collisions
as well, and the study of $d$$+$Au collisions allows us to determine
these effects directly and to disentangle them from the effects
of the hot and dense nuclear medium.

One of the most intriguing discoveries in the early days of the
research program at the Relativistic Heavy Ion Collider 
was the significantly enhanced baryon production
relative to meson production at intermediate transverse momentum
2~GeV/$c$~$<p_T<$~5~GeV/$c$, as evidenced in the large baryon to meson
ratios and the significant differences in the particle
suppression patterns measured by the nuclear modification
factors~\cite{ppg015,ppg026,Adams:2003am}.
Several classes of models were introduced to explain these
differences based on different physical phenomena, such as strong
radial flow~\cite{flow2002,flow2004},
baryon junctions~\cite{Vitev:2001zn,Vitev:2002wh},
and hadronization through
recombination~\cite{Hwa:2002tu,Fries:2003vb,Fries:2003kq,Greco:2003xt,Greco:2003mm,Molnar:2003ff}.

Additionally, the recombination model has been employed to explain
the baryon vs. meson difference in the Cronin enhancement observed
in p+A collisions~\cite{Hwa:2004zd,Hwa:2004yi}.
Traditional explanations of the Cronin enhancement involve the
multiple soft scatterings in the initial state prior to the hard
scattering and subsequent fragmentation of the hard scattered
parton~\cite{Accardi:2002ik}.  This process can naturally explain
the deficit of particle production
at low $p_T$ and enhancement at intermediate $p_T$, but
does not account for the particle species dependence at 
Relativistic-Heavy-Ion-Collider energies~\cite{ppg030}.
Conversely, in the recombination model, the observed enhancement
is attributed to final state effects, i.e. the recombination of
soft partons from the nuclear medium with hard scattered partons
in a jet.  For this reason $d$$+$Au collisions represent an
excellent testing ground for the recombination model, since hot
nuclear matter effects, such as the collective expansion of the
medium, are not expected.
However, recent results in $d$$+$Au collisions at 200~GeV~\cite{ppg152}
and $p$$+$Pb collisions at
5.02~TeV~\cite{CMS:2012qk,Abelev:2012ola,Aad:2012gla} suggest
some collective expansion effects may be present in the most central events.

Measurements of strange particles, such as charged kaons, have
also been an interesting subject in heavy ion collisions.  An
enhancement of strangeness production relative to that in $p$$+$$p$
collisions has been observed at various colliding
energies~\cite{strange1999}.
This strangeness enhancement is a possible signature of
deconfinement, thermalization, and flavor
equilibration~\cite{Rafe:1982,Koch:1986}.
In this scenario, strangeness production is dominated by thermal
gluon fusion.  The measurement of charged kaons in a broad $p_T$
range and in different centrality classes is a significant tool
to further understand the thermalization of the system and the
mechanism of strangeness production.

To address the particle production in both hot and cold QCD matter,
a systematic study of identified particles over a broad $p_T$ range
with a wide selection of centralities  in both Au$+$Au and $d$$+$Au
collisions is required.  In this paper, the spectra, particle
ratios, and nuclear modification factors previously reported by
PHENIX in Au$+$Au~\cite{ppg026} and $d$$+$Au collisions~\cite{ppg030}
are revisited, extending the $p_T$ reach of previous measurements
and significantly improving the statistical precision.
In Section~\ref{s:exp} we discuss the experimental apparatus
and the detector subsystems used in this analysis; in
Section~\ref{s:analysis} we discuss the analysis method,
including event and track selection, and particle identification;
in Section~\ref{s:results} we discuss the results; and in
Section~\ref{s:summary}, we summarize our findings.

\section{Experimental Setup}
\label{s:exp}

The PHENIX experiment is a large, general purpose detector
with a wide variety of detector subsystems ideally suited to
the study of nuclear matter in conditions of extreme
temperature and density.
PHENIX is composed of global event property detectors,
forward and backward rapidity arms (North and South)
dedicated to muon measurements, and two central arm
spectrometers (East and West) at midrapidity covering
pseudorapidity region of $|\eta|<$~0.35 for measurements of
photons, electrons, and charged hadrons.  Detailed
descriptions of the various detector subsystems can be found
in~\cite{Adcox:2003zm}.

\begin{figure}[htb]
\includegraphics[width=1.0\linewidth]{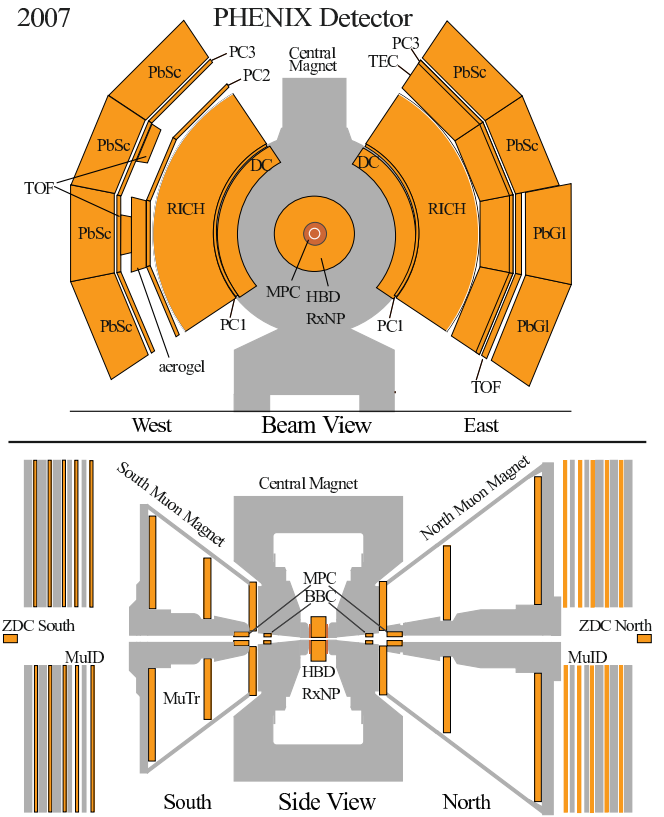}
\caption{(Color online) The PHENIX detector configuration
for the 2007 data taking period.  The 2008 configuration is
very similar.}
\label{fig:phenix}
\end{figure}

Figure~\ref{fig:phenix} gives a schematic diagram of the PHENIX 
detector for the 2007 configuration; the 2008 configuration is very 
similar.  The upper panel shows the central 
spectrometer arms, viewed along the beam line facing North. The lower 
panel shows the two forward rapidity muon arms (North and South) and the 
global detectors.

The analysis presented herein makes use of the beam-beam
counters, the tracking system (drift chamber and two layers
of pad chambers),
the electromagnetic calorimeter,
and the time-of-flight detector in the West arm.

\subsection{Detector subsystems}
\label{s:detectors}

The beam-beam counters (BBC) are used for the minimum-bias trigger,
the centrality definition, the determination of collision vertex
along the beam axis (the $z$-vertex), and the event start time.
The BBCs~\cite{nimbbc,phenixinner} are located at $\pm$~144~cm
from the nominal interaction point.  They cover 2$\pi$ in azimuth
and a pseudorapidity range of 3.0~$<|\eta|<$~3.9.  Each BBC is an
array of 64 identical hexagonal detector elements, with the beam
pipe passing through the center of the array.
Each element is a quartz \v{C}erenkov
radiation counter, and the radiator and photomultiplier tube are
constructed as a single piece.

The PHENIX tracking system is optimized for the high multiplicity
environment of ultra-relativistic heavy ion collisions.  It
comprises drift chambers
(DC)~\cite{nimdc} and pad
chambers (PC1, PC2, and PC3)~\cite{nimpc}.  This
analysis makes use of the first layer of pad chambers (PC1)
and the third layer (PC3).  The DCs have an active volume in the
radial range 2.02~m~$<r<$~2.46~m.  The PC1 is mated directly to
the back of the DC frame in each arm at a radial distance of
2.49~m.  The PC3 is located in each arm at a radial distance of
4.98~m.

The primary particle identification detector used in this analysis
is the time-of-flight detector in the West half of the central
arm spectrometer (TOFW).  The TOFW~\cite{brian,ppg123}
is located at a radial distance
of 4.81~m from the interaction point, having
pseudorapidity coverage of $|\eta|<$ 0.35 and azimuthal coverage
of 22$^\circ$ in two separate sections.  The individual elements
are multigap resistive plate chambers (MRPCs).  Each MRPC has six
230~$\mu$m gas gaps separated by five 550~$\mu$m thick glass plates.
On each side of the outermost glass plates
(1.1~mm thick) are carbon tape electrodes held at +7~kV on one 
side and -7~kV on the other side for a total bias voltage of 
14~kV.
When charged particles traverse the detector, the gas between the
plates is ionized and the image charge is collected at each side
of the chamber on four copper readout strips.
Each strip has dimensions 37 $\times$ 2.8~cm$^{2}$ with a
separation of 0.3~cm between them.
The strips are oriented
lengthwise along the azimuthal direction.  Each strip is read out
from both top and bottom so that the time difference between them
can be used to determine the hit position along the length of the
strip with resolution of order 1~cm.  The TOFW system is composed
of a total of 128 MRPCs, 512 strips, and 1024 readouts.  The total
timing resolution, which includes the uncertainty in the start
time from the BBC, is 84~ps in Au$+$Au collisions~\cite{ppg123}.
In $d$$+$Au
collisions it is 95~ps, where the slightly poorer
resolution is due to the lower resolution of the start time
determination from the BBC.  This is due to the lower multiplicity
in $d$$+$Au collisions and, for the same reason, the $z$-vertex
resolution is also poorer in $d$$+$Au collisions.

\section{Analysis Method}
\label{s:analysis}

\subsection{Event and Track Selection}
\label{s:events_and_tracks}

This paper presents an analysis of Au$+$Au collisions at
$\sqrt{s_{_{NN}}}$~=~200~GeV, collected in 2007, 
and $d$$+$Au collisions at $\sqrt{s_{NN}}$~=~200~GeV, collected 
in 2008.  For each set we select events that pass the minimum-bias trigger,
which is defined as a coincidence between the North and South BBCs.  In Au$+$Au
collisions, this trigger requires two or more photomultiplier tubes firing
in each BBC and measures 92$\pm$3\% of the total inelastic cross section; in
$d$$+$Au collisions, it requires one or more photomultiplier tubes firing in each
BBC and measures 88$\pm$4\% of the total inelastic cross section.  We have an
additional requirement that the collision vertex is within $|z|<$~30~cm of the
nominal origin of the coordinate system.

Centrality selection is performed with the BBCs using the Glauber Monte 
Carlo procedure described in~\cite{Miller:2007ri}, in which the charge in 
each BBC detector is assumed to be proportional to the number of 
participating nucleons $N_{\rm part}$ traveling towards it.  For the 
Au$+$Au system the north and south BBC distributions are summed, but for 
the $d$$+$Au system, only the south (Au-going) side is used.  The BBC 
charge is assumed to follow a negative binomial distribution (NBD) with a 
mean of $N_{\rm part}$ and the remaining NBD parameters determined from a 
$\chi^2$ minimization of the combined Glauber+NBD calculation with respect 
to the data.  The BBC distributions are divided into equal probability 
bins, and the corresponding Glauber distributions are used to calculate 
$N_{\rm part}$ as well as the number of binary nucleon-nucleon collisions 
$N_{\rm coll}$, as shown in Table~\ref{t:centrality}.

\begin{table}[htbp]
    \caption{Values of $\langle N_{\rm coll} \rangle$ and $\langle N_{\rm part} \rangle$
      for Au$+$Au and $d$$+$Au collisions from Glauber model simulations.}\label{t:centrality}
  \begin{center}
    \begin{ruledtabular}
    \begin{tabular}{rrr}
      Centrality & $\langle N_{\rm coll} \rangle$ & $\langle N_{\rm part} \rangle$ \\
      \hline
      Au$+$Au & & \\
       0--10\% & 960.2 $\pm$ 96.1 & 325.8 $\pm$ 3.8 \\
      10--20\% & 609.5 $\pm$ 59.8 & 236.1 $\pm$ 5.5 \\
      20--40\% & 300.8 $\pm$ 29.6 & 141.5 $\pm$ 5.8 \\
      40--60\% &  94.2 $\pm$ 12.0 &  61.6 $\pm$ 5.1 \\
      60--92\% &  14.8 $\pm$  3.0 &  14.7 $\pm$ 2.9 \\
      \hline
      $d$$+$Au & & \\
       0--20\% & 15.1 $\pm$ 1.0 & 15.3 $\pm$ 0.8 \\
      20--40\% & 10.2 $\pm$ 0.7 & 11.1 $\pm$ 0.6 \\
      40--60\% &  6.6 $\pm$ 0.4 &  7.8 $\pm$ 0.4 \\
      60--88\% &  3.1 $\pm$ 0.2 &  4.3 $\pm$ 0.2 \\
      0--100\% &  7.6 $\pm$ 0.4 &  8.5 $\pm$ 0.4 \\
    \end{tabular}
    \end{ruledtabular}
  \end{center}
\end{table}

Charged track reconstruction in the DC is based on a combinatorial
Hough transform, which gives the angle in the main bend plane
($r$-$\phi$) and thus $p_T$.  The PC1 is used to determine the hit
position in the longitudinal ($z$) direction.
Only tracks with valid information in both the DC and PC1 are used in
this analysis.  Tracks in DC/PC1 are projected to the outer detectors,
such as PC3 and TOFW, and matched to hits in those detectors with
the minimum distance between the projection and the hit position.  The
distribution of differences between hits and projections is
approximately Gaussian, with an additional underlying background
caused by random associations.  Only tracks with a difference of less
than two standard deviations in both the azimuthal and longitudinal
directions in both the PC3 and the TOFW are selected, so as to
minimize background contamination.
In the Au$+$Au data for
$p_T>$~5.0~GeV/$c$, an additional background isolation cut is applied.
For these tracks we require $E/p_T>$~0.2, where $E$ is the energy
deposited in the electromagnetic calorimeter.  This cut removes low
$p_T$ particles that are falsely reconstructed as high $p_T$ tracks.

\subsection{Particle Identification}
\label{s:PID}

Charged particle identification (PID) is performed by simultaneous
measurement of the momentum, time-of-flight, and path-length.  These
quantities are used to determine the mass of the candidate track based
on the following relationship:

\begin{equation}
m^2 = \frac{p^2}{c^2}\Bigl(\frac{t^2c^2}{L^2}-1\Bigr),
\end{equation}
where $m$ is the mass, $p$ is the momentum, $c$ is the speed of light,
$t$ is the time of flight, and $L$ is the path-length.  The $m^2$
distributions are approximately Gaussian.  The standard deviation of
the distribution $\sigma_{m^2}$ can be parametrized as a function of
momentum as follows:

\begin{eqnarray}
{\sigma^2_{m^2}} &=& 
\frac{\sigma_{\alpha}^2}{K_{1}^2}\bigl(4m^4p^2\bigr)
+ \frac{\sigma_{ms}^2}{K_{1}^2}\Bigl(4m^4 \Bigl(1+\frac{m^2}{p^2} \Bigr) \Bigl) \nonumber \\
&+& \frac{\sigma_{t}^2c^2}{L^2}\bigl(4p^2 \bigl(m^2 + p^2 \bigr) \bigr),
\label{eq:pid}
\end{eqnarray}
where $\sigma_{m^2}$ is the standard deviation of the $m^2$ distribution,
$m$ denotes the physical mass of the particle and thus the square root
of the centroid of the $m^2$ distribution, $\sigma_{\alpha}$ is the
angular resolution of the drift chamber, $\sigma_{ms}$ is the multiple
scattering term, $\sigma_{t}$ is the total timing resolution, and $K_1$
is the magnetic field integral constant.  The magnetic field integral
constant depends on the magnetic field configuration.  The PHENIX magnet
system for the central arms comprises two coils.  The two coils can
be run together, opposed, or with the inner coil off.
During the 2007 Au$+$Au data taking, the coils were run opposed, while for
the 2008 $d$$+$Au data taking, the coils were run together.
Running the coils opposed produces zero magnetic field in the region
between the beam-pipe and the inner coil of the magnet.  This is needed
for the analysis of the dielectron continuum using the hadron blind
detector, which is the innermost detector during this operational
period and can be seen (labeled as HBD) in Fig.~\ref{fig:phenix}.
%
%
The PID parameters are presented in Table~\ref{t:pidpars}.

\begin{table}[htbp]
    \caption{Parameters for the PID function defined in Equation~\ref{eq:pid}.}\label{t:pidpars}
  \begin{center}
    \begin{ruledtabular}
      \begin{tabular}{llrr}
        Parameter & & 2007 Au$+$Au & 2008 $d$$+$Au \\
        \hline
        $\sigma_{\alpha}$ & (mrad)               & 0.896   & 1.050 \\
        $\sigma_{ms}$     & (mrad$\cdot$GeV/$c$) & 0.992   & 1.000 \\
        $\sigma_{t}$      & (ps)                 & 0.084   & 0.095 \\
        $K_1$             & (mrad$\cdot$GeV/$c$) & 75.0    & 104.0 \\
      \end{tabular}
    \end{ruledtabular}
  \end{center}
\end{table}

\begin{figure}[htbp]
\includegraphics[width=1.0\linewidth]{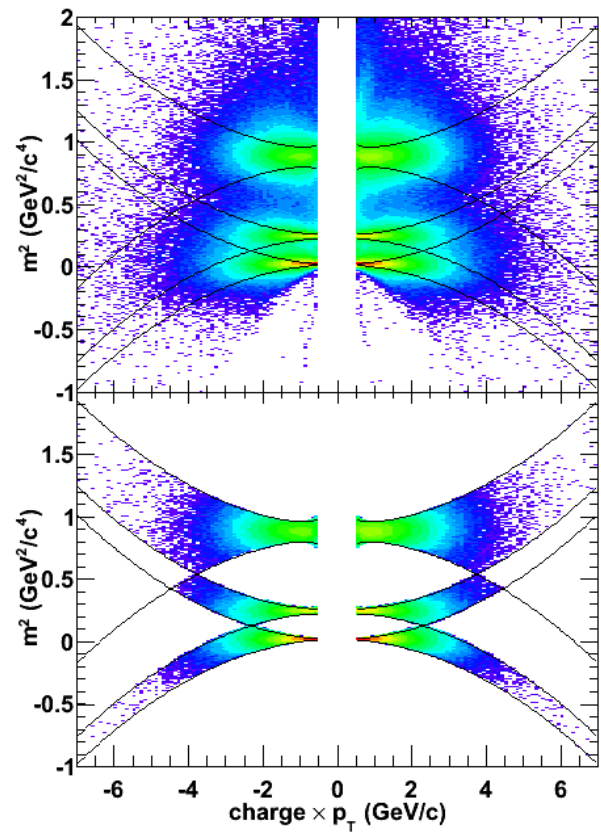}
\caption{(Color online) The particle identification method,
$m^2$ vs charge $\times$ $p_T$ for 2007 Au$+$Au data; the 2008
$d$$+$Au data are very similar. The solid black lines indicate the two
standard deviation PID bands used for the cuts.  The upper
panel shows the bands superimposed on the entire $m^2$
distribution, the lower panel shows the distribution after
the cuts have been applied.} \label{fig:pidbands}
\end{figure}

To select candidate tracks of a particular particle species, the $m^2$ is
required to be within two standard deviations of the mean for the selected
particles species and outside two standard deviations of the mean for the
other particle species.
Below the regions where the PID bands intersect, the PID contamination is
negligible.  Above it, contamination is from one side of the tail of the
distribution beyond 2 standard deviations.  For a Gaussian distribution
this is 2.25\%. Therefore the contamination can be estimated based on the
ratio of the respective yields multiplied by this value.
The $m^2$ distributions may have slightly nonGaussian tails and
therefore the PID contamination may be slightly higher.  In any event,
we estimate the PID purity to be better than 90\% for all particle
species at all $p_T$, in all centrality classes, and in both collision
systems.
%
Figure~\ref{fig:pidbands} shows the $m^2$ vs the $p_T$ multiplied by the
charge for the 2007 Au$+$Au data; the 2008 $d$$+$Au data are very similar.  
The 2$\sigma$ PID bands are superimposed as solid black lines.  The upper
panel shows the entire $m^2$ distribution with all the
track selection cuts applied but none of the PID cuts.  The lower panel
shows the same distribution but with the PID cuts also applied.

\subsection{Corrections to the Raw Data}
\label{s:corr}

In order to obtain the true invariant yield, the raw spectrum needs to be
corrected for a variety of factors.  Various types of simulations are
performed to determine these corrections.  To correct for geometrical
acceptance, analysis cuts, particle interactions with detector materials,
and in-flight decays (for pions and kaons), we use single particle Monte
Carlo (MC) simulations.  For these simulations, single particles are
generated using a random generator, with flat distributions in rapidity,
azimuth, and $p_T$.  The random particles are then run through a {\sc geant}
simulation of PHENIX to determine the interactions of the single particles
with the detector subsystems and support structures.  Next, all the detector
response information is fed through the usual PHENIX reconstruction software
to produce simulated tracks.  Finally, these simulated tracks are analyzed in
the exact same way as tracks from the real data in order to determine the
corrections.  The total correction factor, $F_C(p_T)$, is given by the
following relation:

\begin{equation}
F_C(p_T) = \frac{dN_{{\rm output}}/dp_T}{dN_{{\rm input}}/dp_T} 
= \epsilon_{{\rm acceptance}}\epsilon_{{\rm efficiency}}\epsilon_{{\rm cuts}}.
\end{equation}

To correct for the detector occupancy effect, which is most significant in
the TOFW, we run embedding simulations, where a track from single particle
MC simulations is embedded into a real event, and the occupancy correction
is determined from the relative efficiencies of reconstructing the single
track in isolation and in the event.  This correction is the largest in the
most central Au$+$Au collisions where the multiplicity is the highest and
therefore the occupancy effect is the strongest.  In the most peripheral
Au$+$Au collisions the multiplicity is low enough that there is essentially
no effect.  The same is true in $d$$+$Au collisions, where no correction is
applied.

For the $d$$+$Au system we apply a correction for the underlying event correlation
that exists between produced particles measured in the central arms and particles
at forward angles that satisfy the BBC interaction trigger~\cite{Adler:2004eh}.
This correlation produces both a trigger bias, in which events  satisfying the
trigger are biased towards higher multiplicities, and a bin shift, in which
nominally peripheral events are shifted to higher centrality bins, thereby
depleting the more peripheral bins.  We correct for these
effects using the Glauber Monte Carlo combined with central particle yields
measured in $p$$+$$p$ collisions.  Using this same framework, we also generate
a correction factor to convert the minimum-bias sample (0--88\%) into one with
zero-bias (0--100\%).

Table~\ref{t:centcorr} shows the centrality dependent corrections for each
collision species.  The occupancy is represented as an efficiency, while the bias
factor is represented as a multiplicative correction.

\begin{table}[htbp]
    \caption{Summary of centrality dependent corrections.}\label{t:centcorr}
  \begin{center}
    \begin{ruledtabular}
      \begin{tabular}{rr}
        Au$+$Au & Occupancy Dependent Efficiency \\
        \hline
         0--10\% & 0.542 \\
        10--20\% & 0.653 \\
        20--40\% & 0.783 \\
        40--60\% & 0.904 \\
        60--92\% & 0.964 \\
	\hline
	$d$$+$Au & Bias Factor Correction \\
	\hline
	 0--20\% & 0.94 \\
	20--40\% & 1.00 \\
	40--60\% & 1.03 \\
	60--88\% & 1.03 \\
	0--100\% & 0.89 \\
      \end{tabular}
    \end{ruledtabular}
  \end{center}
\end{table}

The proton and antiproton spectra are additionally corrected for the
feed-down from weak decays of  hyperons into protons.  We use single
particle MC simulations of the $\Lambda$ baryon and apply the analysis
cuts used for the protons to determine the percentage of $\Lambda$
baryons that decay into protons that pass our proton selection cuts.
This is used to determine the percentage of the total proton
sample that likely comes from hyperon decays, which is called the
feed-down fraction.

The feed-down fraction is only dependent on the $\Lambda/p$ ratio
and not explicitly on the $\Lambda$ spectrum itself.
For the Au$+$Au data we take the $\Lambda/p$ ratio to be 0.89 and the
$\bar{\Lambda}/\bar{p}$ ratio to be 0.95, as measured in Au$+$Au collisions
at $\sqrt{s_{_{NN}}}$~=~130~GeV~\cite{Adcox:2002au}.  These are very similar
to (and well within the systematic uncertainties of) the values of 0.91
for $\Lambda/p$ and 0.94 for $\bar{\Lambda}/\bar{p}$ obtained for Au$+$Au
collisions at $\sqrt{s_{_{NN}}}$~=~200~GeV by examining the $dN/dy$ values
for  $\Lambda$ and $\bar{\Lambda}$~\cite{Adams:2006ke} and $p$ and
$\bar{p}$~\cite{ppg026}.
We note however that the $\Lambda$ data reported in Ref.~\cite{Adcox:2002au}
are inclusive while the data $\Lambda$ reported in Ref.~\cite{Adams:2006ke}
are corrected for feed-down from other hyperons.

For the $d$$+$Au data, we take the $\Lambda/p$ ratio to be 0.85 and the
$\bar{\Lambda}/\bar{p}$ ratio to be 0.99.  There are no published data on
$\Lambda/p$ and $\bar{\Lambda}/\bar{p}$ ratios nor $\Lambda$ and
$\bar{\Lambda}$ spectra or yields in $d$$+$Au collisions that could be used
to estimate these ratios.  Instead, we use the
$(\Lambda+\bar{\Lambda})/(p+\bar{p})$ ratio measured in $p$+$\bar{p}$
collisions~\cite{Ansorge:1989ba}, the $\bar{p}/{p}$ ratio measured in
$d$$+$Au collisions~\cite{ppg030}, and an estimate of the
$\bar{\Lambda}/\Lambda$ ratio in $d$$+$Au collisions based on measurements in
Au$+$Au~\cite{Hippolyte:2003yf} and $p$$+$$p$ collisions~\cite{Billmeier:2004cs}.

The spectral shape of the $\Lambda$ ($\bar{\Lambda}$) is assumed to follow
the $p$ ($\bar{p}$) spectrum with $m_T$ scaling.  We also take the ratio to
be independent of centrality.  In fact, a small centrality dependence can be
seen when examining the integrated yields in
Refs.~\cite{ppg026}~and~\cite{Adams:2006ke},
although the different centralities are consistent within the systematic
uncertainties.

All measurements of $\Lambda$ implicitly include the $\Sigma^0$, which decays
electromagnetically with 100\% branching ratio to the $\Lambda$ and a photon.
We do not correct for feed-down from the charged $\Sigma$ states, nor for the
$\Xi$ and $\Omega$ multistrange baryon states.  Because these corrections are
smaller and have large uncertainties, they are included in the overall systematic
uncertainty estimates.  For the charged $\Sigma$ states we are concerned only with
the $\Sigma^+$,  which decays to a proton with a 0.5 branching ratio.  The data
from 200~GeV $p$+$\bar{p}$ collisions show a ratio
$(\Sigma^++\Sigma^-+\bar{\Sigma}^++\bar{\Sigma}^-)/(\Lambda+\bar{\Lambda})$
of 0.50~$\pm$~0.18(syst) Ref.~\cite{Ansorge:1989ba}.  
Assuming $\bar{\Sigma}/\Sigma~\approx~\bar{\Lambda}/\Lambda$
and $\Sigma^-/\Sigma^+~\approx$~1, we can estimate
$\Sigma^+/\Lambda$ to be 0.25.  The ratios for
$(\Xi^0+\bar{\Xi}^0)/(\Lambda+\bar{\Lambda})$
and
$(\Xi^-+\bar{\Xi}^+)/(\Lambda+\bar{\Lambda})$
in $p$+$\bar{p}$ collisions both have a value of 0.065 with very large
($\approx$100\%) systematic uncertainties.  Assuming
$\bar{\Xi}/\Xi~\approx~\bar{\Lambda}/\Lambda$ one can estimate the
$(\Xi^0+\Xi^-)/\Lambda$ ratio to be roughly 0.13, which is consistent with the
total feed-down correction of 15\% reported for $\Lambda$ in Ref.~\cite{Adams:2006ke}.
The latter  also includes feed-down from $\Omega$, though the contribution is small.

The systematic
uncertainty on the feed-down fraction is estimated to be 25\%, which
is due primarily to the uncertainty in the $\Lambda/p$ ratio.
We also note that, in the $p_T$ region of interest to this paper,
2~GeV/$c$~$<~p_T~<$~5~GeV/$c$, the feed-down fraction is of order
10\%, so a 25\% change in the $\Lambda/p$ ratio
produces only a roughly 2.5\% change in the proton spectrum.

\begin{figure}[htbp]
\includegraphics[width=0.97\linewidth]{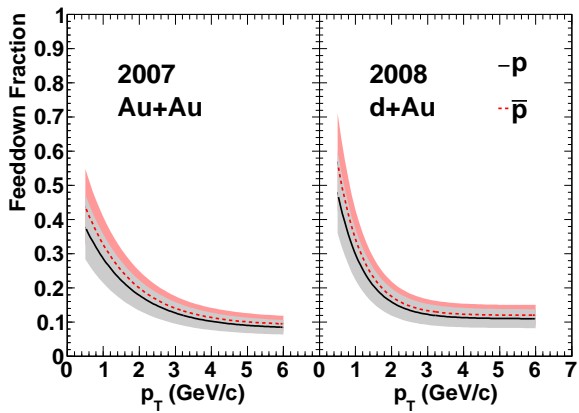}
\caption{(Color online) Feed-down fraction as a function of $p_T$
for protons (sold black lines) and antiprotons (dotted red lines)
for the 2007 Au$+$Au data (left panel) and the 2008 $d$$+$Au data
(right panel).  The shaded bands indicate the systematic
uncertainty of 25\%.
} \label{fig:feedplot}
\end{figure}

Figure~\ref{fig:feedplot} shows the feed-down fraction as a function
of $p_T$ for protons as solid black lines and antiprotons as dotted
red lines for the 2007 Au$+$Au data (left panel) and 2008 $d$$+$Au data
(right panel).  As mentioned above, the magnetic field
configuration is different for 2007 and 2008, which is the reason
for the different feed-down fractions between the two data sets.

\subsection{Estimation of Systematic Uncertainties}
\label{s:sys}

The different types of systematic uncertainty are categorized as follows:
type A, point-to-point uncorrelated in $p_T$, where the points can move up
or down independently of each other; type B, point-to-point correlated
in $p_T$, where the points can move together changing the shape of the
curve; and type C, an overall normalization uncertainty in which all points
move up or down by the same factor.  The corrections for analysis cuts,
including acceptance, track selection, and PID, are predominantly type B.
The normalization corrections for effects such as detector occupancy and
efficiency are type C.  The uncertainties are assessed for each cut by
redoing the analysis with the cut varied and then determining the
difference.  The cuts are varied in exactly the same way for the analysis
of both the experimental and the simulated data.  This process is repeated
for all the analysis cuts and the differences are summed in quadrature to
determine the final uncertainty.  This is done for both the spectra and for
the ratios.  The uncertainties are examined in each centrality class for
both Au$+$Au and $d$$+$Au collisions.  In Au$+$Au collisions, they are found to be
quite similar in all centrality classes.  In $d$$+$Au
collisions they are found to be negligibly different for all centrality
classes.
When taking ratios of the various particles, all of the type C uncertainties
and some of the type B uncertainties cancel.  For antiparticle to particle
ratios, the uncertainty from acceptance does not cancel at all, and the
uncertainty from track selection and PID mostly cancel.  For other particle
ratios, $K/\pi$ and $p/\pi$, the uncertainty  from acceptance and track
selection mostly cancel, and the uncertainty from PID does not cancel at all.
In this analysis, we find the remaining systematic uncertainty on each of the
particle ratios is roughly 5\% for all $p_T$.
For the nuclear modification factor $R_{CP}$, which compares two different
centrality bins of the same particle species in the same collision system,
all the type B systematic uncertainties cancel almost completely.  There is
an uncertainty of about 2\% based on small variations of the
track matching and PID distributions as a function of centrality.  The type C
uncertainties are completely uncorrelated and added in quadrature.  For the
nuclear modification factors $R_{\rm AA}$ and $R_{dA}$, the $p$$+$$p$ reference
data were collected during a different operational period and using different
detector subsystems, therefore none of the systematic uncertainties cancel.
A summary of the type B systematic uncertainties for the spectra is given in
Table~\ref{t:specsys}.

\begin{table}[h]
\caption{Summary of systematic uncertainties from acceptance, 
track selection, and PID of invariant yield of each particle 
species.}\label{t:specsys}
\begin{ruledtabular}
\begin{tabular}{lrrrrrr}
 & $\pi^{+}$ & $\pi^{-}$ & $K^{+}$ & $K^{-}$ & $p$ & $\bar{p}$ \\
\hline
2007 Au$+$Au       &      &      &      &      &      &      \\
$p_T<$ 3 GeV/$c$ &  9\% &  9\% & 11\% & 11\% & 10\% & 10\% \\
$p_T>$ 3 GeV/$c$ & 10\% & 10\% & 11\% & 11\% & 11\% & 11\% \\
$p_T>$ 5 GeV/$c$ & 14\% & 14\% & -    & -    & 14\% & 14\% \\
\hline
2008 $d$$+$Au        &      &      &      &      &      &      \\
$p_T<$ 3 GeV/$c$ &  8\% &  8\% & 13\% & 13\% &  9\% &  9\% \\
$p_T>$ 3 GeV/$c$ &  9\% &  9\% & 13\% & 13\% & 11\% & 11\% \\
\end{tabular}
\end{ruledtabular}
\end{table}

The type C uncertainties are from the centrality dependent corrections for the
spectra and from the uncertainty on the Glauber model calculations for the
nuclear modification factors.  The uncertainty on the occupancy correction for
Au$+$Au collisions is roughly 10\%, and therefore the uncertainty on the yield
varies from 5\% for the most central to less than 1\% for the most peripheral.
The uncertainties for the Glauber values for $N_{\rm coll}$ are much larger and
therefore dominate the uncertainty in the nuclear modification factors.
The uncertainty for the bias factors for $d$$+$Au collisions varies from about 1\%
for the most central bin to about 5\% for the most peripheral bin.  The bias
factors are determined in the same Glauber model analysis as the $N_{\rm coll}$
and $N_{\rm part}$ values, and therefore the uncertainties are correlated.  The
uncertainty on the ratio of bias factors and $N_{\rm coll}$ values used to
determine the nuclear modification factors varies from about 3\% for the most
central to about 8\% for the most peripheral.

\section{Results and Discussion}
\label{s:results}

\subsection{Invariant Yield as a Function of Transverse Momentum $p_T$}

\begin{figure*}[htbp]
\centering
\includegraphics[width=0.99\linewidth]{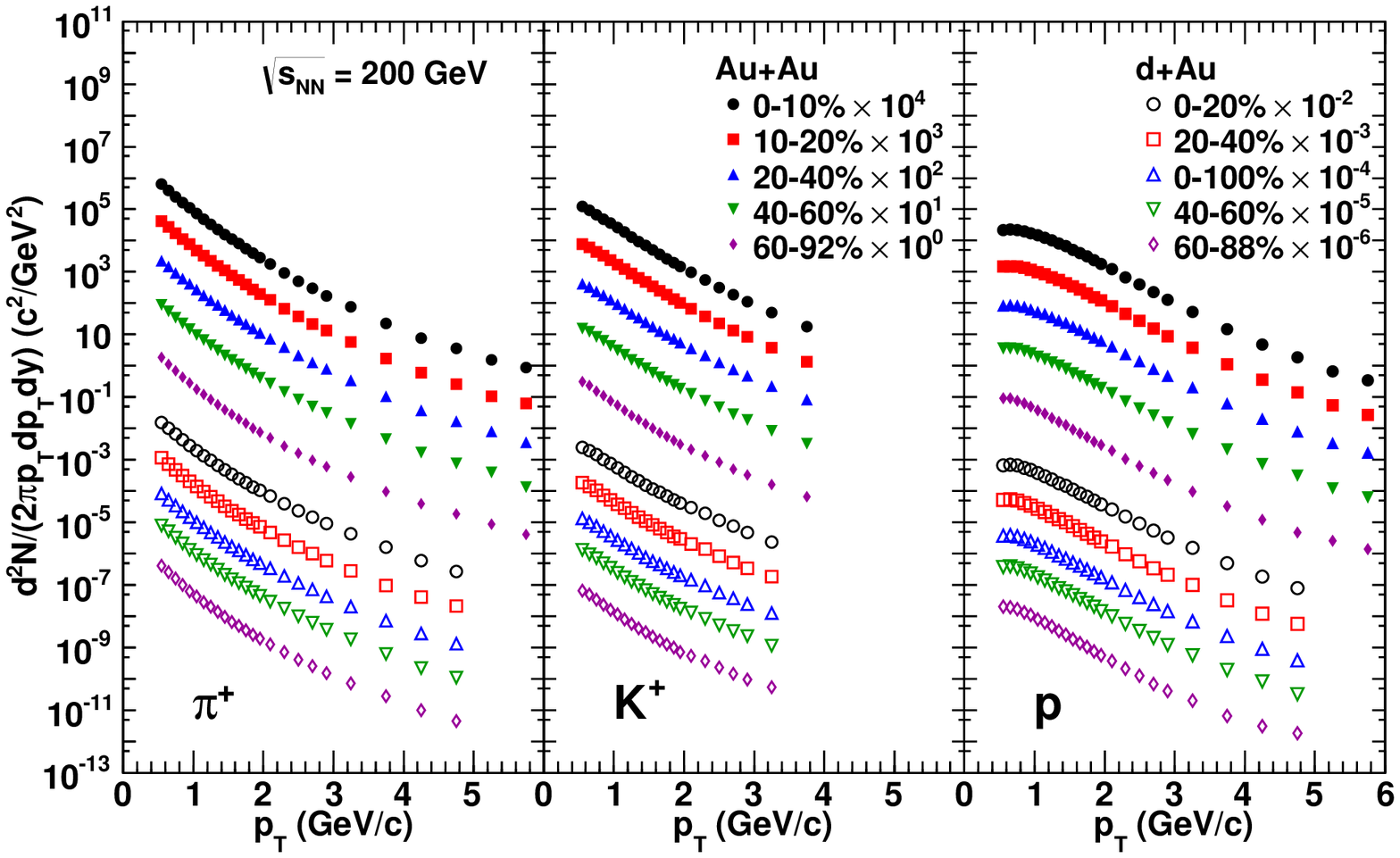}
\includegraphics[width=0.99\linewidth]{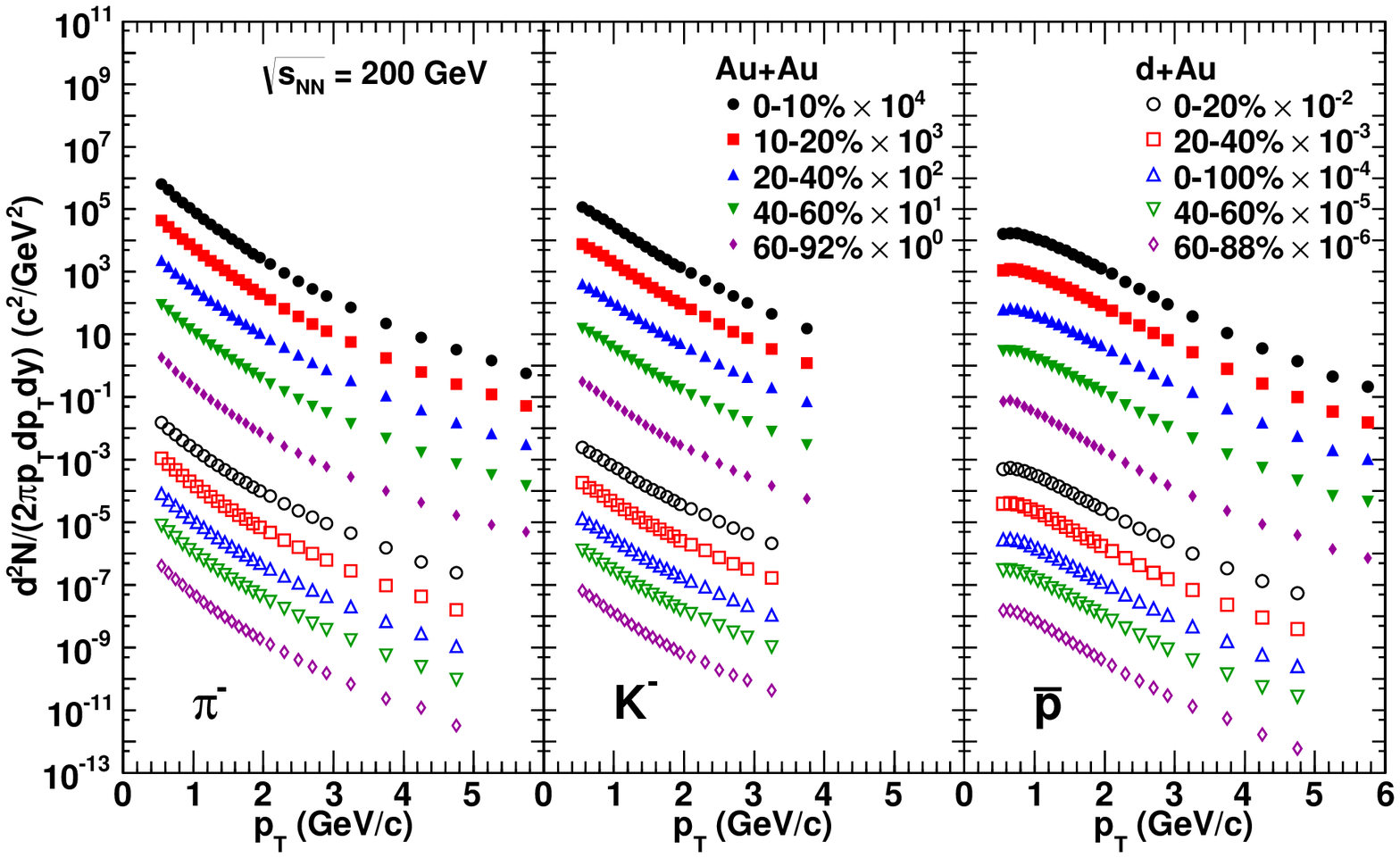}
\caption{(Color online) Invariant yield of $\pi^{\pm}$, $K^{\pm}$,
and $p$ and $\bar{p}$ as a function of
$p_T$ in Au$+$Au and $d$$+$Au collisions.  The yields are scaled by the
arbitrary factors indicated in the legend, keeping collisions
species grouped together.}\label{fig:spectra3}
\end{figure*}

The main result of this study is the measurement of the invariant yield of
pions, kaons, and protons as a function of $p_T$ in different centrality
classes.  The centrality classes studied in the Au$+$Au measurement are 0--10\%
(the most central), 10--20\%, 20--40\%, 40--60\%, and 60--92\% (the most
peripheral).  For the $d$$+$Au measurement, the centrality classes are 0--20\%
(the most central), 20--40\%, 40--60\%, 60--88\% (the most peripheral), and
0--100\%.  From these quantities, all other observables are derived, such as
particle ratios and nuclear modification factors.  
Figure~\ref{fig:spectra3} shows the invariant yields of positive pions,
positive kaons, and protons (upper left, middle, and right, respectively),
and negative pions, negative kaons, and antiprotons (lower left, middle, and
right, respectively).  The yields are scaled by arbitrary factors indicated
in the legend for the sake of clarity and
to keep the collision species grouped together.

\subsection{Particle Ratios as a Function of Transverse Momentum}

\begin{figure*}[htbp]
\centering
\includegraphics[width=0.9\linewidth]{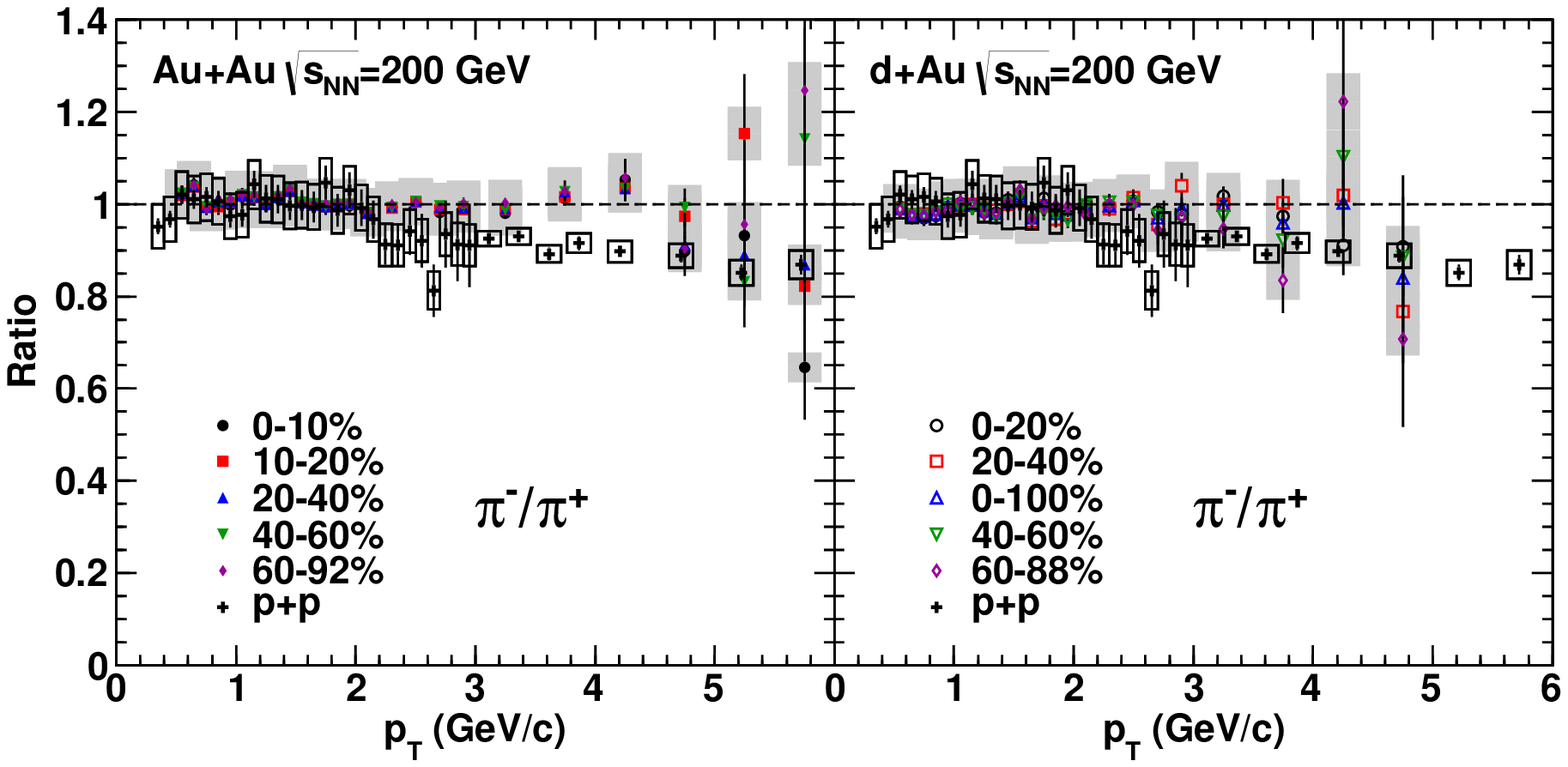}
\includegraphics[width=0.9\linewidth]{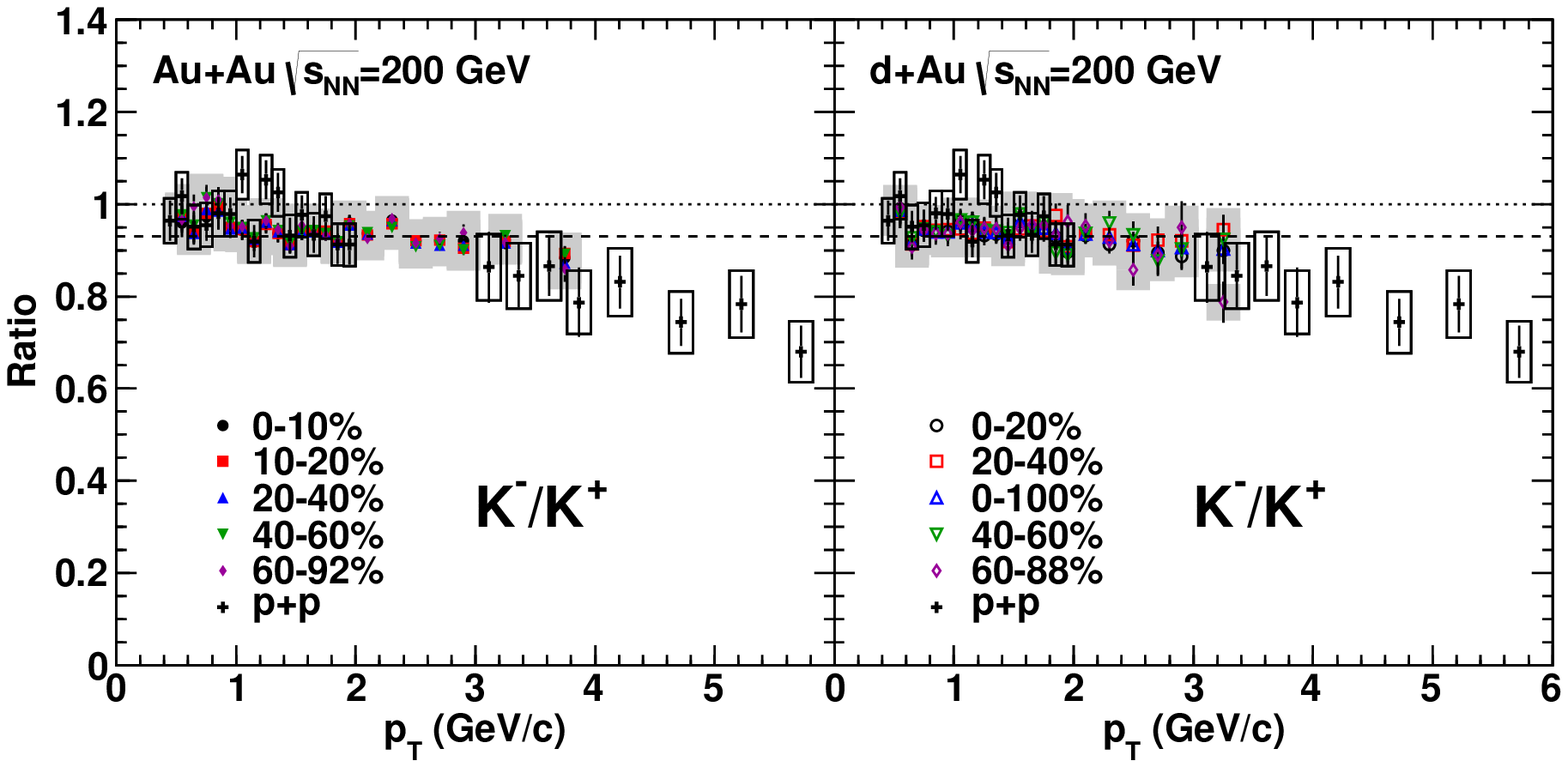}
\includegraphics[width=0.9\linewidth]{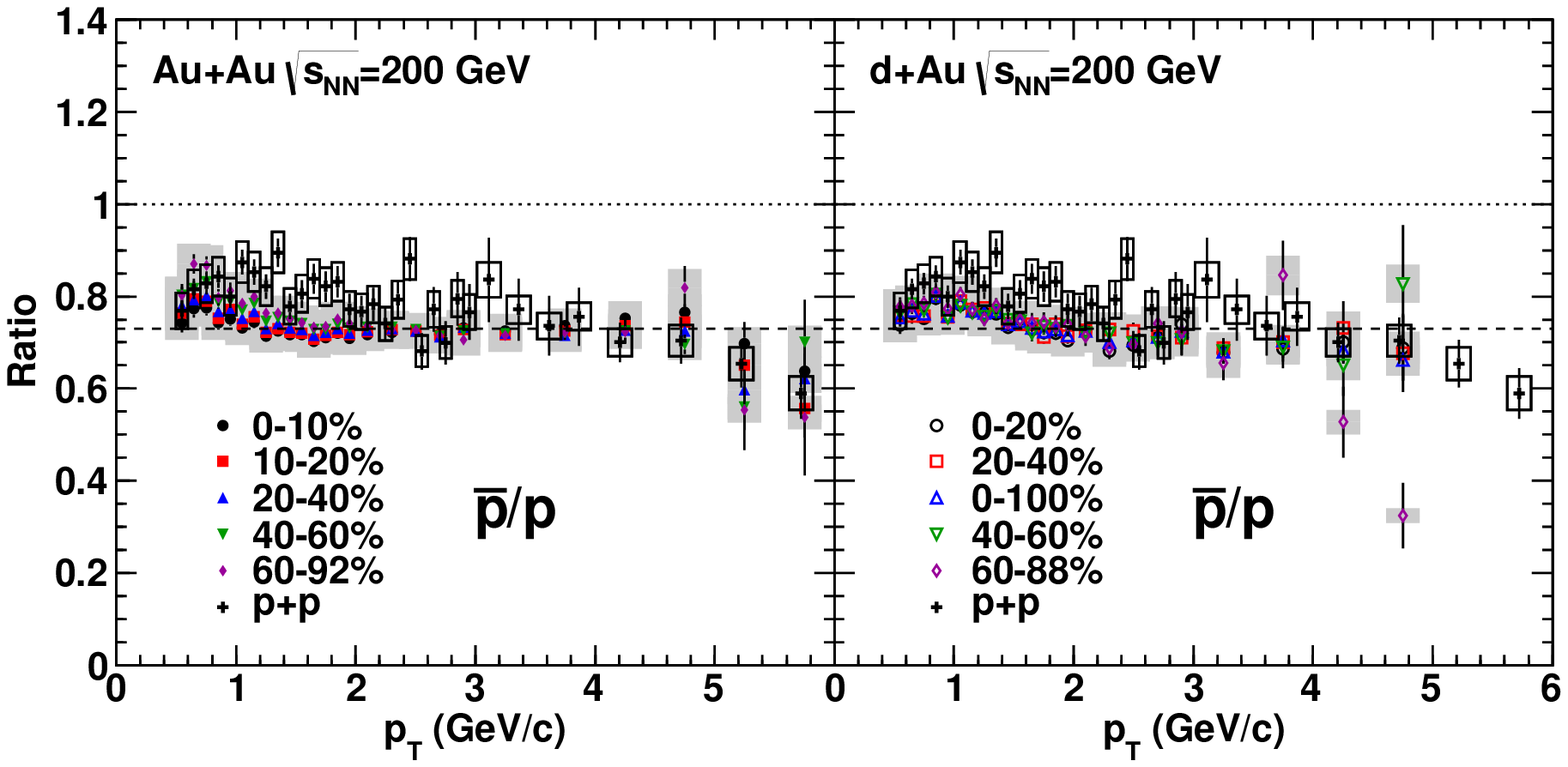}
\caption{(Color online) Ratio of invariant yield of $\pi^{-}/\pi^{+}$
(top), $K^{-}/K^{+}$ (middle), and $\bar{p}/p$ (bottom) as a function
of $p_T$ in Au$+$Au (left panels) and $d$$+$Au collisions (right panels) in
each centrality bin.  Dashed lines are drawn as a visual aid with
values of 1.0 for $\pi^{-}/\pi^{+}$, 0.93 for $K^{-}/K^{+}$, and 0.73
for $\bar{p}/p$.  These values are taken from Ref.~\cite{ppg026}.
Shown as a reference are data from $p$$+$$p$ collisions from Ref.~\cite{ppg101}
for $p_T<$~3~GeV/$c$ and from Ref.~\cite{Agakishiev:2011dc} for $p_T>$~3~GeV/$c$.
}\label{fig:homrat}
\end{figure*}

One of the simpler classes of derived quantities is the antiparticle to
particle ratio.  In the present analysis those ratios are $\pi^{-}/\pi^{+}$,
$K^{-}/K^{+}$, and $\bar{p}/p$, which are plotted as a function of $p_T$ in
the upper, middle,  and lower panels of Fig.~\ref{fig:homrat}, respectively.
In each panel, the Au$+$Au data are on the left and the $d$$+$Au data are on the
right.  Drawn as a visual aid are dashed black lines with value 1.0 for the
pions, 0.93 for the kaons, and 0.73 for the protons; the values for kaons and
protons are chosen from the reported $p_T$ integrated values from
Ref.~\cite{ppg026}.
Shown as a reference are data from $p$$+$$p$ collisions from Ref.~\cite{ppg101}
for $p_T<$~3~GeV/$c$ and from Ref.~\cite{Agakishiev:2011dc} for $p_T>$~3~GeV/$c$.
Remarkably, all the ratios are essentially independent of
both $p_T$ and centrality.
Based on simple arguments about isospin
conservation and the basics of the parton distribution functions and
fragmentation functions, one would expect each of the antiparticle to particle
ratios to vary as a
function of $p_T$ as discussed in Ref.~\cite{Wang:1998bha}.
Indeed, these ratios have a
significant $p_T$ dependence in $p$$+$$p$ collisions at
midrapidity, and the agreement with theory depends
significantly on the fragmentation functions used~\cite{Agakishiev:2011dc}.
However, the $p_T$ range needed to observe the decrease in these ratios in
$p$$+$$p$ collisions is quite large.  As seen in Fig.~\ref{fig:homrat}, the
$p_T$ dependence of the ratios in $p$$+$$p$ is small and may be consistent with 
the ratios in $d$$+$Au and Au$+$Au.

\begin{figure}[htp]
\centering
\includegraphics[width=0.948\linewidth]{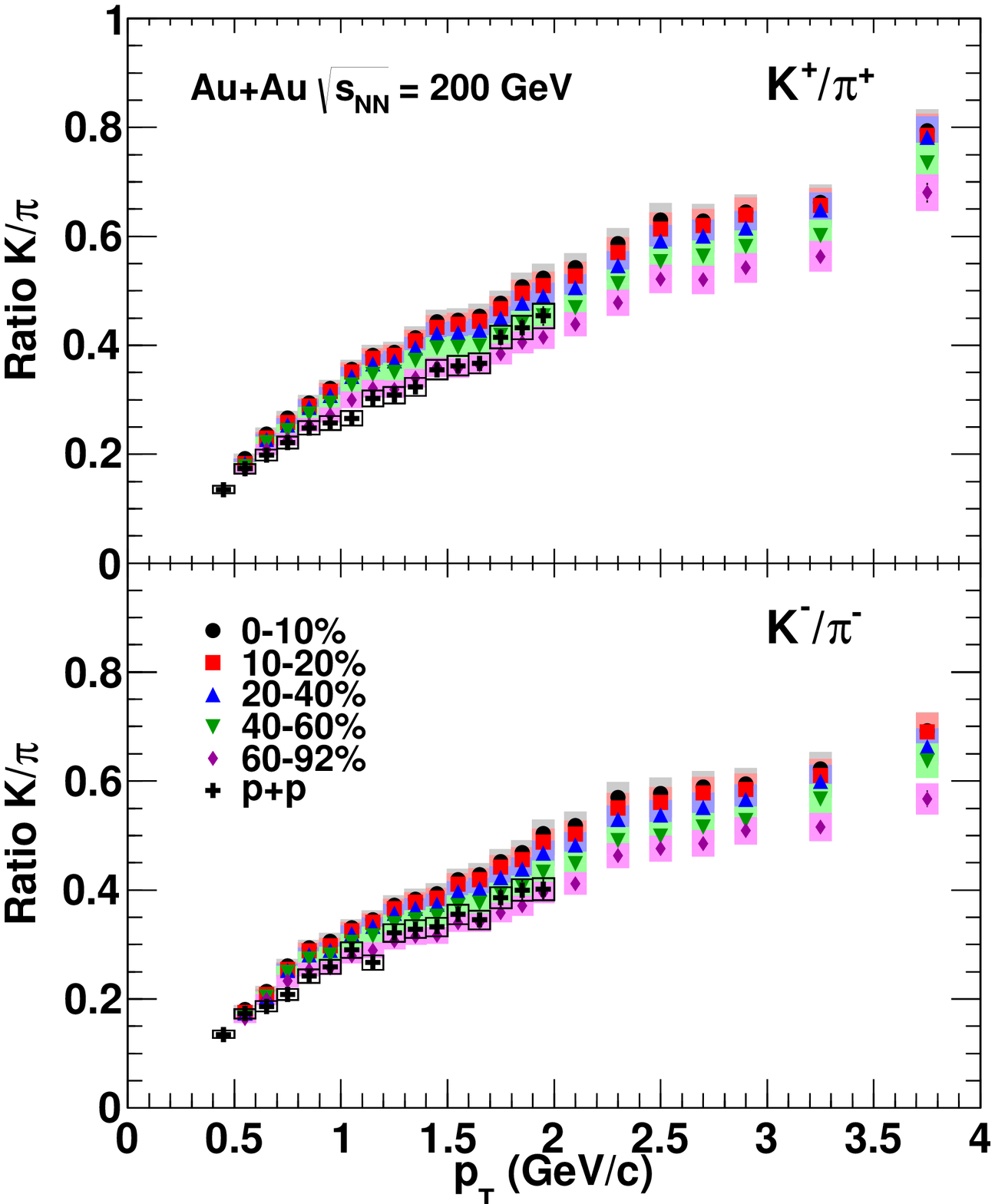}
\caption{(Color online) Ratio of invariant yield of positive kaons
to positive pions (upper panel) and negative kaons to negative pions
(lower panel) as a function of $p_T$ in Au$+$Au collisions in
the centrality bins marked in the legend.  Data for $p$$+$$p$
collisions~\protect\cite{ppg101} are shown as
a reference.}\label{fig:kpiratio_auau}
\end{figure}

\begin{figure}[htp]
\centering
\includegraphics[width=0.948\linewidth]{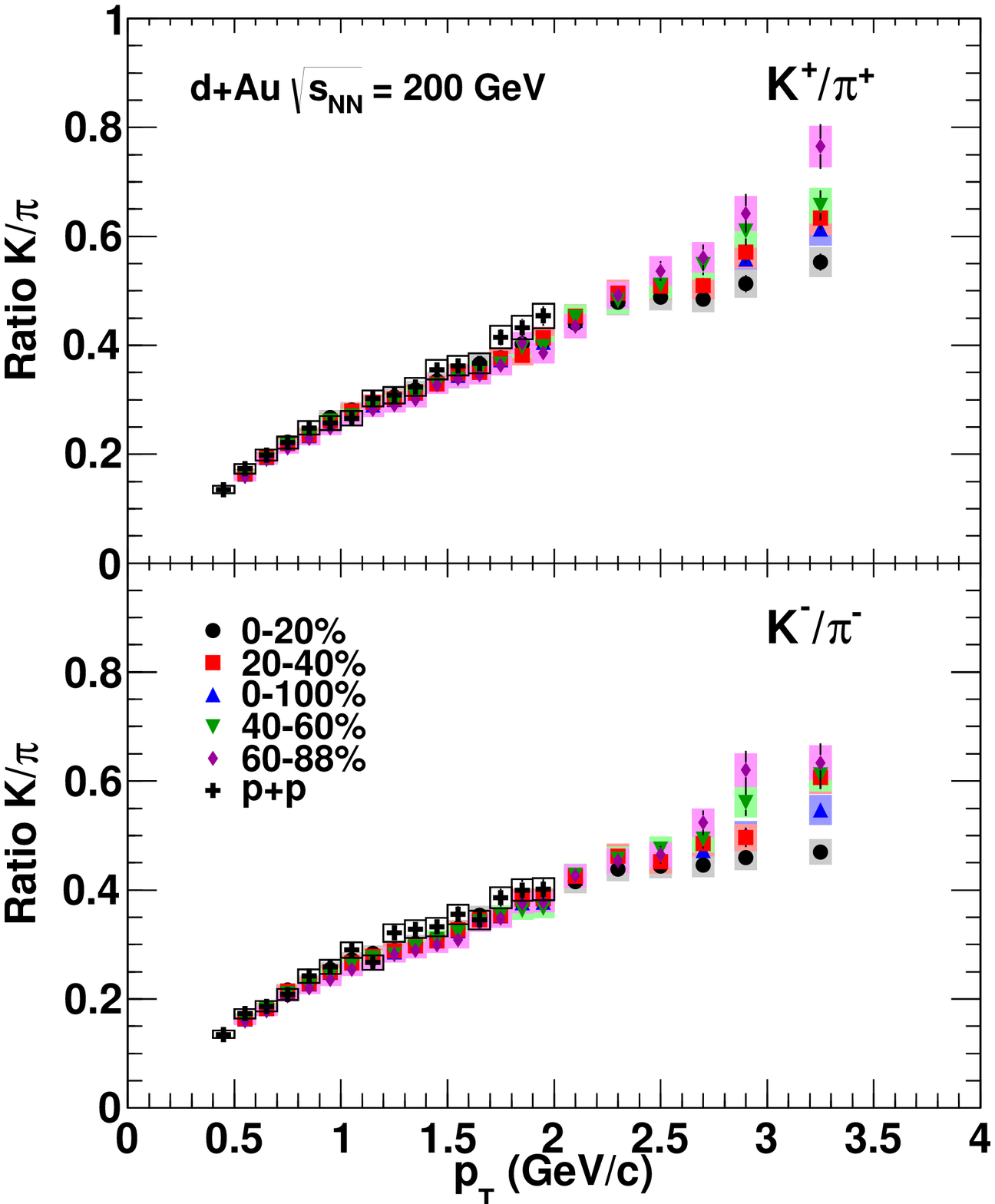}
\caption{(Color online) Ratio of invariant yield of positive kaons
to positive pions (upper panel) and negative kaons to negative pions
(lower panel) as a function of $p_T$ in $d$$+$Au collisions in
the centrality bins marked in the legend.  Data for $p$$+$$p$
collisions~\protect\cite{ppg101} are shown as 
a reference.}\label{fig:kpiratio_dau}
\end{figure}

Figures~\ref{fig:kpiratio_auau} and~\ref{fig:kpiratio_dau} show the
kaon to pion ratios as a function of $p_T$ ($K^{+}/\pi^{+}$ on the top,
$K^{-}/\pi^{-}$ on the bottom) in Au$+$Au and $d$$+$Au collisions, respectively.
The ratios in Au$+$Au collisions show a significant increase with increasing
$p_T$ and a small increase as the collisions become more central.
The enhancement of
the integrated $K/\pi$ ratio in more central collisions is attributed to
strangeness equilibration in various thermal
models~\cite{Kaneta:2004zr,Cleymans:2004pp}, which is reflected in the
differential ratio. 
However, the differential ratio may include additional
information about the differences in the fragmentation functions and/or the
phase space distribution functions used in the recombination models.
As
discussed in a previous PHENIX publication~\cite{ppg096}, the strangeness
enhancement present in the hot and dense nuclear medium has an effect on
certain recombination models~\cite{Hwa:2006vb}.  These recombination models
involve the recombination of partons in dissimilar momentum space, meaning
that a shower parton from a jet can recombine with a thermal parton in the
medium.  The thermal component of thermal+shower recombination is more dominant
at higher $p_T$ for strange hadrons (like kaons) than it is for nonstrange
hadrons (like pions), leading to an enhancement of the ratio that increases
with $p_T$.
This increasing enhancement
manifests as the ratio rising more quickly in Au$+$Au collisions compared
with $p$$+$$p$ collisions, which is seen in Figs.~\ref{fig:kpiratio_auau} 
and~\ref{fig:kpiratio_dau}.
At sufficiently high $p_T$ where the shower component begins to dominate
for both strange and nonstrange particles, this ratio is expected to turn
over and begin to decrease.  However, this turnover point, if it
exists, is beyond the $p_T$ reach available for kaons in this study.

The $K/\pi$ ratios in $d$$+$Au collisions are essentially identical for all
centrality classes, which may indicate that the mechanism for strangeness
production in $d$$+$Au collisions is the same for all centrality classes.
However, we also note that the various $d$$+$Au centrality classes span a
relatively small range of $N_{\rm part}$.  Therefore, if the strangeness
enhancement is only weakly dependent on $N_{\rm part}$, the variation of
$N_{\rm part}$ in the $d$$+$Au centrality classes may not be large enough for
an effect to be observed.

\begin{figure}[htp]
\centering
\includegraphics[width=0.948\linewidth]{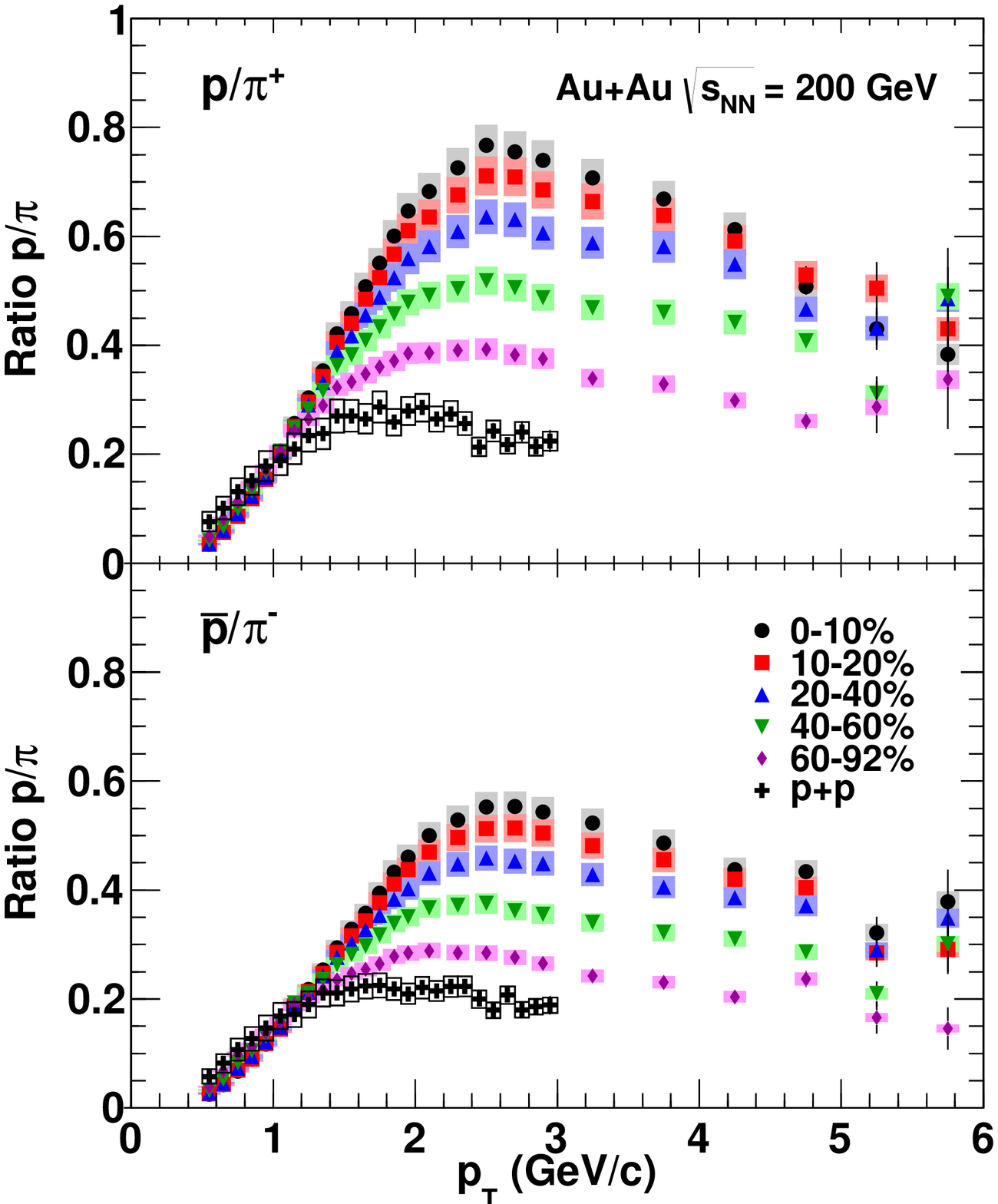}
\caption{(Color online) Ratio of invariant yield of protons to
positive pions (upper panel) and antiprotons to negative pions
(lower panel) as a function of $p_T$ in Au$+$Au collisions in
the centrality bins marked in the legend.  Data for $p$$+$$p$
collisions~\cite{ppg101} are
shown as a reference.}\label{fig:ppiratio_auau}
\end{figure}

\begin{figure}[htp]
\centering
\includegraphics[width=0.948\linewidth]{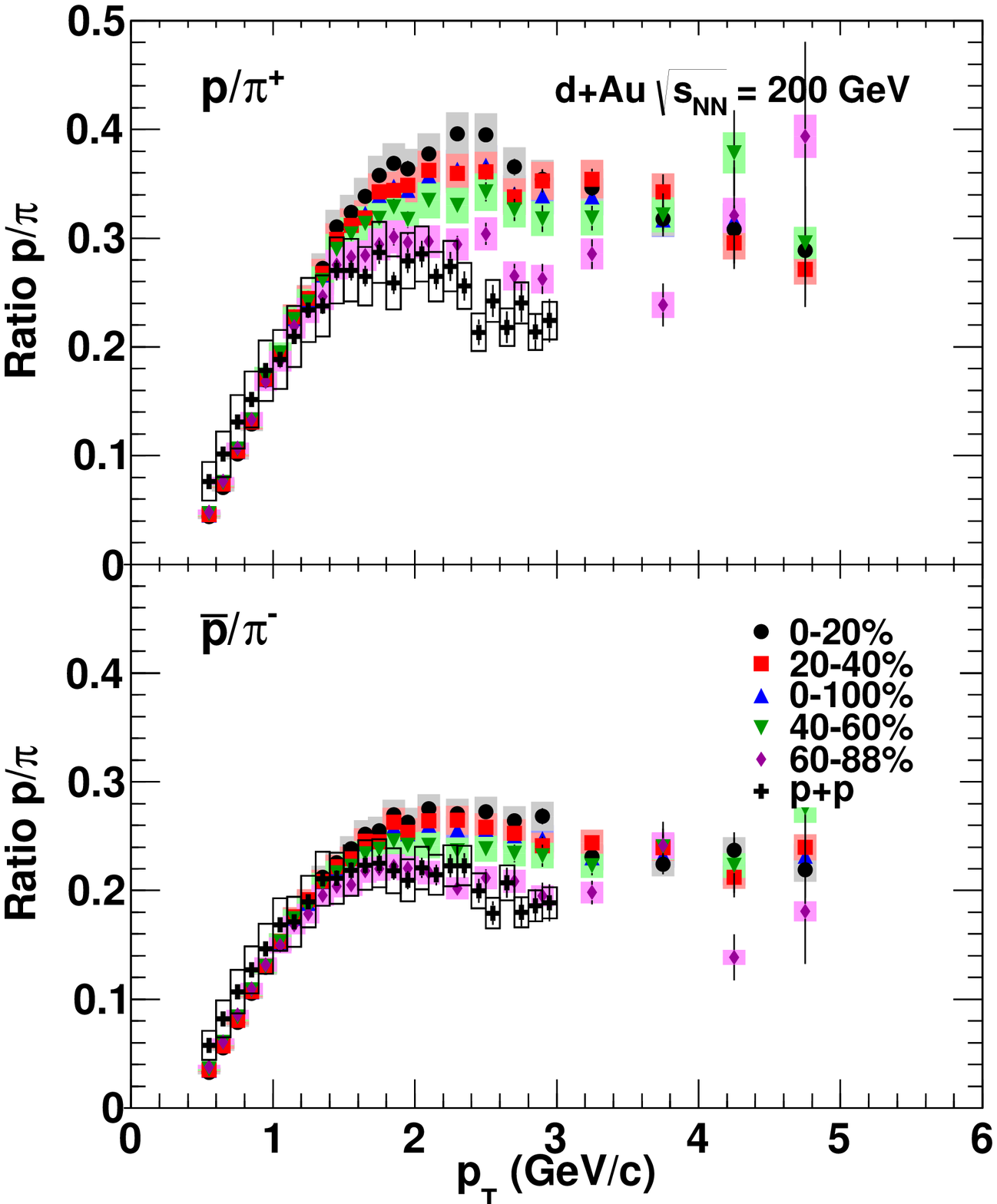}
\caption{(Color online) Ratio of invariant yield of protons to
positive pions (upper panel) and antiprotons to negative pions
(lower panel) as a function of $p_T$ in $d$$+$Au collisions in
the centrality bins marked in the legend.  Data for $p$$+$$p$
collisions~\cite{ppg101} are
shown as a reference.}\label{fig:ppiratio_dau}
\end{figure}

\begin{figure*}[htbp]
\centering
\includegraphics[width=0.99\linewidth]{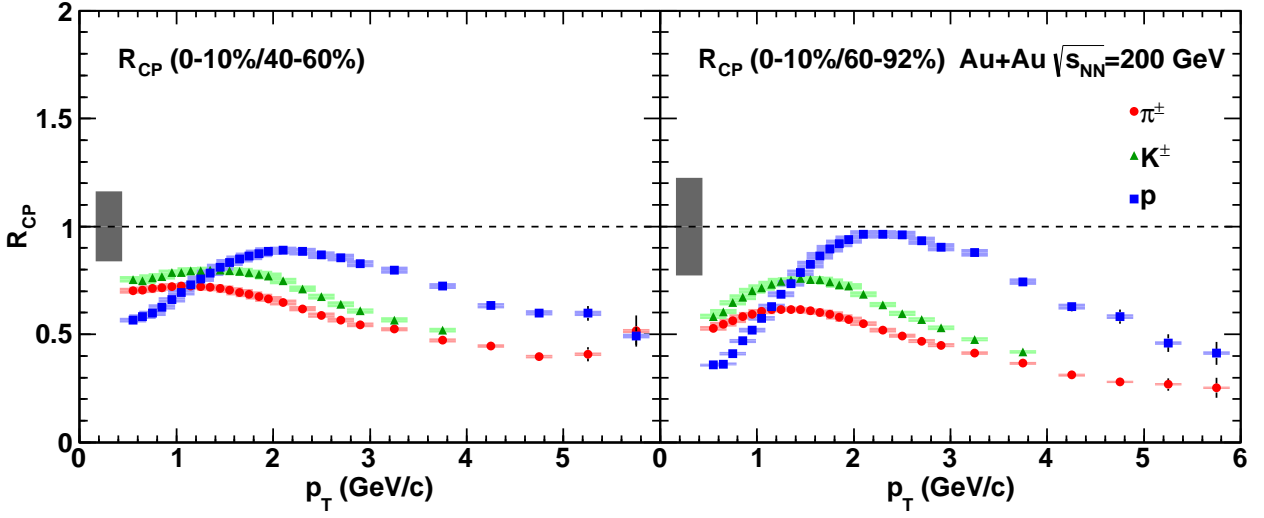}
\caption{(Color online) $R_{CP}$ for 0--10\%/40--60\%
(left panel) and 0--10\%/60--92\% (right panel) as a
function of $p_T$ for charge averaged pions, kaons, and
protons.  A dashed black line is drawn at unity as a visual aid,
indicating nonmodification.  The shaded gray boxes
indicate the associated uncertainty on $N_{\rm coll}$ from
the Glauber model calculations.}\label{fig:rcp_auau}
\end{figure*}

\begin{figure*}[htbp]
\includegraphics[width=0.99\linewidth]{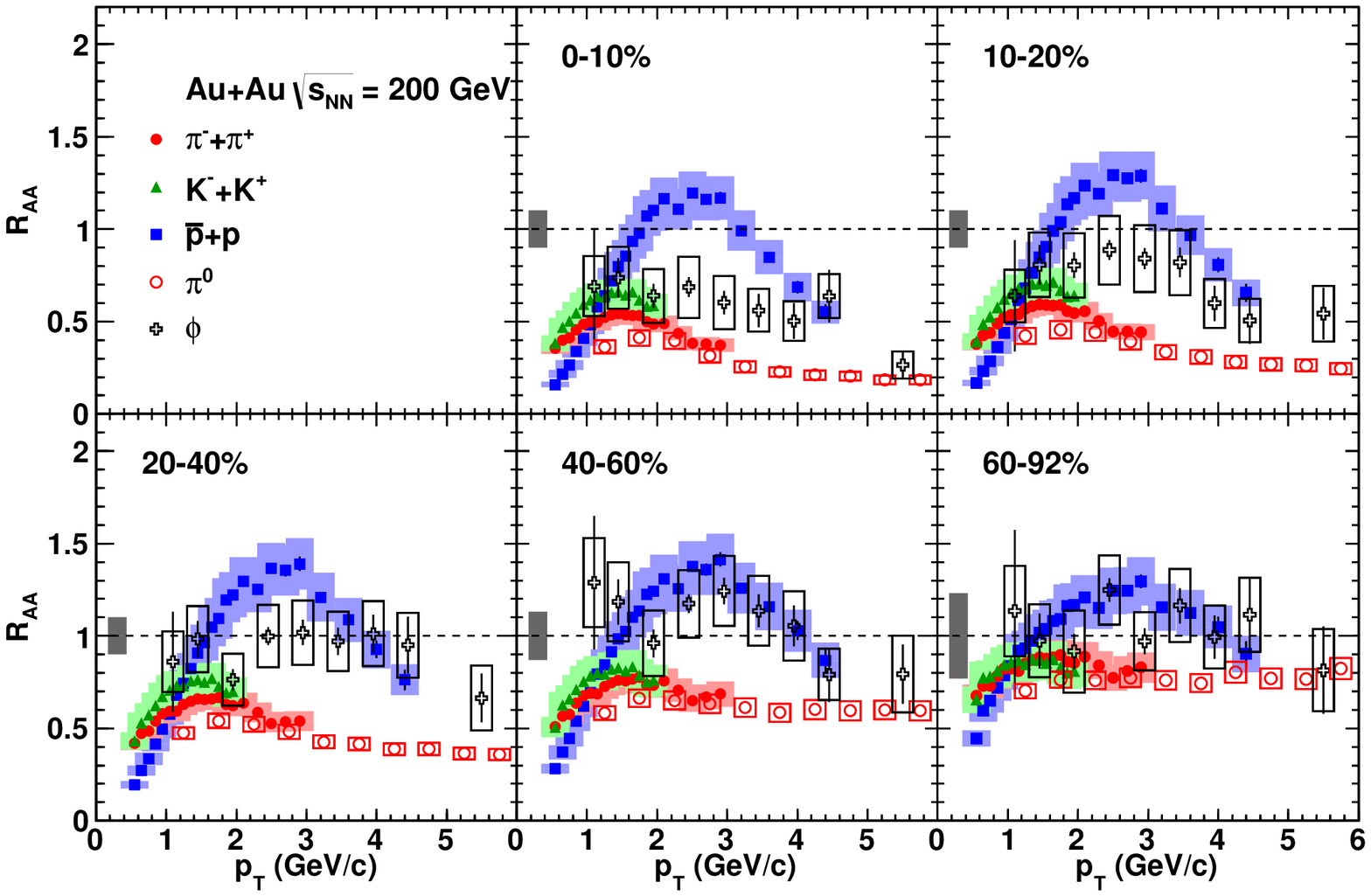}
\caption{(Color online) Nuclear modification factor $R_{\rm AA}$
as a function of $p_T$ in different centrality classes of
charge averaged pions, kaons, and protons,
$\pi^0$~\cite{ppg080}, and $\phi$~\cite{ppg096}.  A dashed
black line is drawn at unity as a visual aid, indicating
nonmodification.  The shaded gray boxes indicate the
associated uncertainty on $N_{\rm coll}$ from the Glauber
model calculations.}\label{fig:raa_auau}
\end{figure*}

Figures~\ref{fig:ppiratio_auau} and~\ref{fig:ppiratio_dau} show the
proton to pion ratios as a function of $p_T$ ($p/\pi^{+}$ on the top,
$\bar{p}/\pi^{-}$ on the bottom) in Au$+$Au and $d$$+$Au collisions, respectively.
Note that for Fig.~\ref{fig:ppiratio_dau} the vertical scale is different.
The ratios in central Au$+$Au collisions show a strong enhancement over the
values in $p$$+$$p$ collisions.  This is conjectured to be attributed to the
parton recombination mechanism of hadronization, which gives rise to a
significant enhancement of baryon yields relative to meson yields in heavy
ion collisions~\cite{Hwa:2002tu,Fries:2003kq,Greco:2003mm}.  The $p/\pi$
ratios in the other centralities in Au$+$Au collisions show a clear and
consistent trend with decreasing enhancement as the collisions become more
peripheral.  In $d$$+$Au collisions there is a similar trend.  The $p/\pi$ ratio
in the most central $d$$+$Au collisions appears consistent with the ratio in
the most peripheral Au$+$Au collisions.  Additionally, the $p/\pi$ ratio is
enhanced over $p$$+$$p$ collisions for each centrality class in $d$$+$Au
collisions except for the most peripheral.

\begin{figure*}[hbtp]
\includegraphics[width=0.99\linewidth]{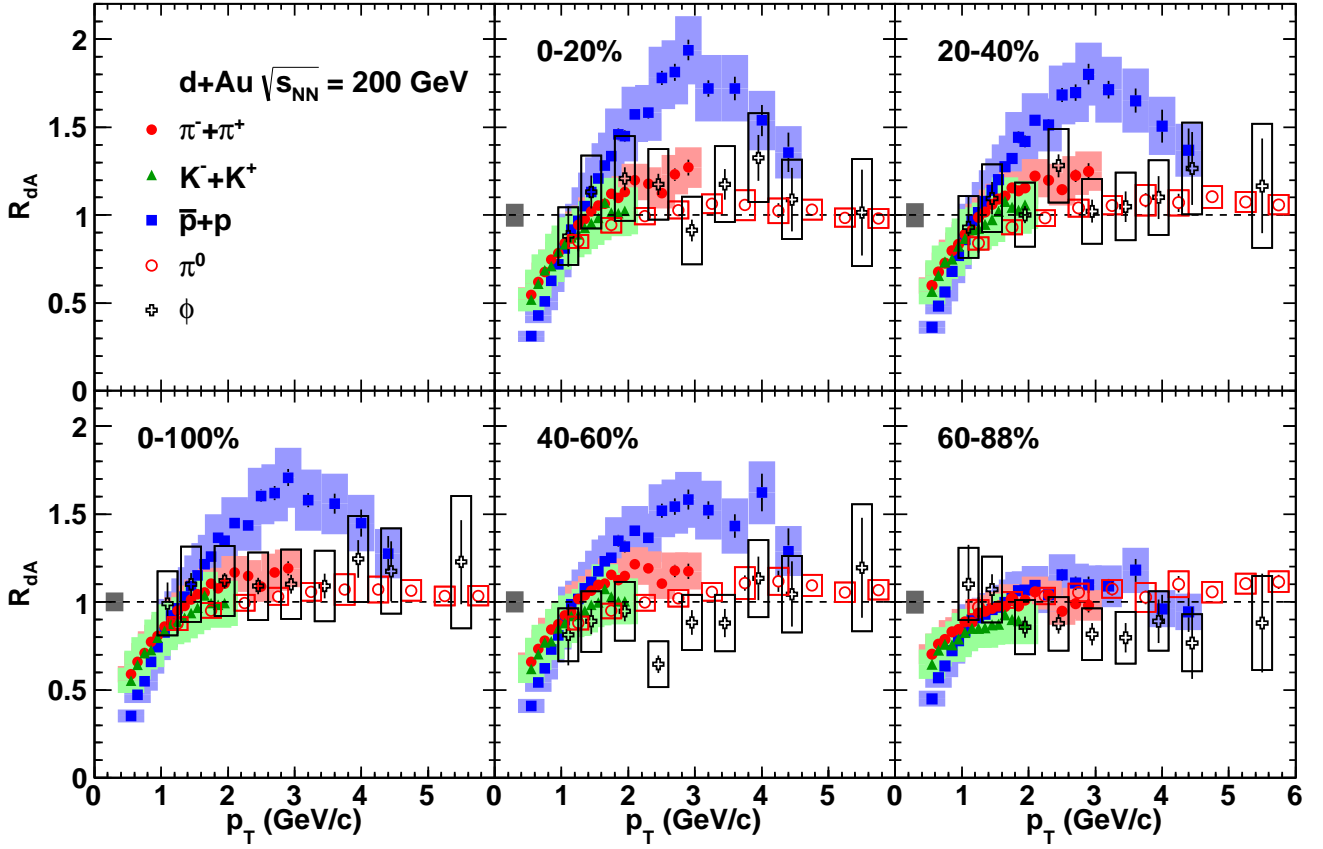}
\caption{(Color online) Nuclear modification factor $R_{dA}$
as a function of $p_T$ in different centrality classes of
charge averaged pions, kaons, and protons,
$\pi^0$~\cite{ppg044}, and $\phi$~\cite{ppg096}.
A dashed black line is drawn at unity as a visual aid,
indicating nonmodification.  The shaded gray boxes
indicate the associated uncertainty on $N_{\rm coll}$ from the
Glauber model calculations.}\label{fig:rda_dau}
\end{figure*}

\subsection{Nuclear Modification Factors as a Function of Transverse Momentum}

To measure the modification of the spectrum of produced particles
in heavy ion collisions relative to the spectrum in $p$$+$$p$ collisions,
nuclear modification factors are employed.  The nuclear modification
factor $R_{\rm AA}$ is defined as the yield in Au$+$Au collisions divided by
the yield in $p$$+$$p$ collisions, normalized by the number of binary
nucleon+nucleon collisions $N_{\rm coll}$ as determined from the Glauber
model.  The nuclear modification factor $R_{CP}$ is defined as the
yield in central Au$+$Au collisions divided by the yield in peripheral
Au$+$Au collisions, normalized to the respective numbers of binary
nucleon+nucleon collisions.  These can be expressed mathematically as:

\begin{equation}
R_{\rm AA} = \frac{(dN/dp_T)^{\rm Au+Au}}
{N_{\rm coll}^{\rm Au+Au}(dN/dp_T)^{p+p}},
\end{equation}
\begin{equation}
R_{CP} = \frac{(dN/dp_T)^{central}}{(dN/dp_T)^{peripheral}}
\frac{N_{\rm coll}^{peripheral}}{N_{\rm coll}^{central}}.
\end{equation} 
Figure~\ref{fig:rcp_auau} shows $R_{CP}$ for 0--10\%/40--60\% (left panel)
and 0--10\%/60--92\% (right panel) as a function of $p_T$ for charge
averaged pions, kaons, and protons.  Both pions and kaons exhibit a
suppression pattern at all values of $p_T$.  The kaons exhibit less
suppression than the pions, indicating the additional role of strangeness
enhancement in the particle production mechanism.  The observed enhancement
of kaons relative to pions appears to be lower for the 0--10\%/40--60\% as
compared to the 0--10\%/60--92\%, suggesting a centrality dependence of
the strangeness enhancement, as seen in the $K/\pi$ ratios discussed above.
The protons on the other hand exhibit quite different behavior, rising to
a value very close to unity, indicating nonsuppression, around 2--3 GeV/$c$
in $p_T$.  At higher values of $p_T$ the proton $R_{CP}$ falls off slowly,
beginning to approach the pion $R_{CP}$ at the highest values of $p_T$
available.  The proton $R_{CP}$ shown here is consistent within the
systematic  uncertainties with the proton $R_{CP}$ reported by
STAR~\cite{Abelev:2006jr}.

Figure~\ref{fig:raa_auau} shows $R_{\rm AA}$ as a function of $p_T$ in different
centrality classes for charge averaged pions, kaons, and protons, as well for
$\pi^0$~\cite{ppg080} and $\phi$~\cite{ppg096}.  We use previously published
PHENIX data on identified hadrons in $p$$+$$p$ collisions~\cite{ppg101} to
evaluate the $R_{\rm AA}$.  The $R_{\rm AA}$ data are limited in $p_T$ reach by the
$p$$+$$p$ data.  As with the $R_{CP}$, the pions and kaons exhibit a suppression
pattern in the $R_{\rm AA}$.  Additionally, a significant and monotonic centrality
dependence is observed, with the suppression decreasing as the collisions
become more peripheral.  This is consistent with what is seen for neutral
pions~\cite{ppg054,ppg080}.  The proton $R_{\rm AA}$ shows no suppression in the
intermediate $p_T$ region and in fact reaches a maximum value above unity
between 2--3~GeV/$c$.  For $p_T>$~3~GeV/$c$, the proton $R_{\rm AA}$ values decrease
and a suppression pattern emerges.  The proton $R_{CP}$ decreases more slowly
than the central proton $R_{\rm AA}$, which is simply because of the still
considerable modification in the peripheral bins.  The trend appears to
be that the proton $R_{CP}$ and $R_{\rm AA}$ decrease steadily while the pion
$R_{CP}$ and $R_{\rm AA}$ hold steady, suggesting that these values for pions
and protons may eventually merge.  The proton $R_{\rm AA}$ for the 0--10\%
centrality bin shown here exhibits reasonable qualitative agreement with the
$K+p$~$R_{\rm AA}$ for 0--12\% centrality reported by STAR~\cite{Agakishiev:2011dc}.

While the centrality dependence of the $R_{\rm AA}$ for the pions and kaons is
strong, it is quite weak for the protons and the different centralities are
consistent within the systematic uncertainties.  This is consistent with the
strong centrality dependence in the $p/\pi$ ratios discussed above.  The $\phi$
meson $R_{\rm AA}$ values are close to the values for kaons, and significantly
lower than the values for the protons, even though the $\phi$ is much heavier
than the kaon and it has roughly the same mass as the proton.  This strongly
suggests a baryon vs.  meson dynamic, as opposed to a simple mass dependence,
as would be the case for radial flow developed during the hadronic phase.

\begin{figure}[bhp]
\includegraphics[width=0.948\linewidth]{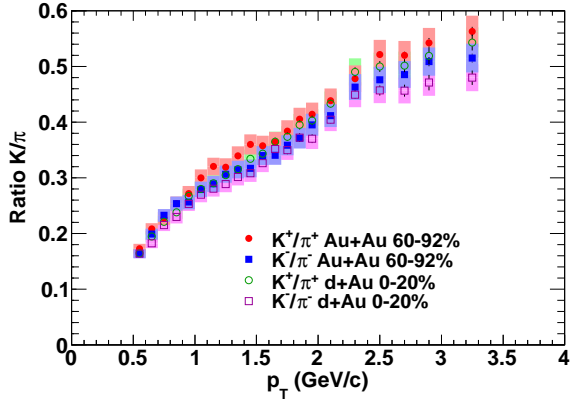}
\caption{(Color online) Ratio of $K^{+}/\pi^{+}$ and $K^{-}/\pi^{-}$ as a
function of $p_T$ in peripheral Au$+$Au and central $d$$+$Au collisions plotted
together.}
\label{fig:kpiratio_aud}
\end{figure}

\begin{figure}[bhp]
\includegraphics[width=0.948\linewidth]{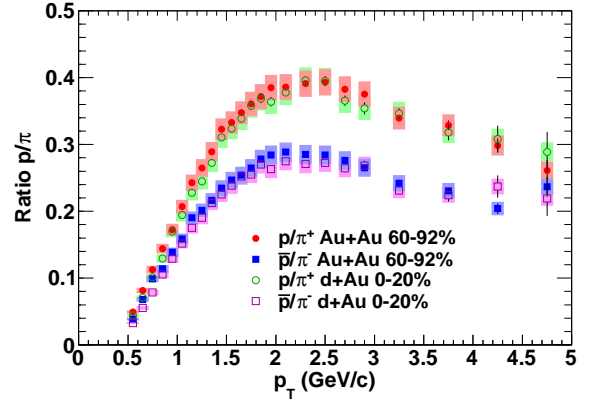}
\caption{(Color online) Ratio of $p/\pi^{+}$ and $\bar{p}/\pi^{-}$ as a
function of $p_T$ in peripheral Au$+$Au and central $d$$+$Au collisions plotted
together.}\label{fig:ppiratio_aud}
\end{figure}

The nuclear modification factor for $d$$+$Au
collisions, $R_{dA}$, is defined in a similar way as $R_{\rm AA}$ by

\begin{equation}
R_{dA} = \frac{(dN/dp_T)^{d+{\rm Au+Au}}}
{N_{\rm coll}^{d+{\rm Au}}(dN/dp_T)^{p+p}}.
\end{equation}
Figure~\ref{fig:rda_dau} shows $R_{dA}$ as a function of $p_T$ in
different centrality classes for charged averaged particles.  As with the
$R_{\rm AA}$, we use previously published PHENIX data on identified hadrons
in $p$$+$$p$ collisions~\cite{ppg101} to evaluate the $R_{dA}$.  The
$p$$+$$p$ data limit the $p_T$ reach of the $R_{dA}$.
The charged pion exhibits a small modification above $p_T$ of 1.0~GeV/$c$
and is consistent with nonmodification within the systematic uncertainties.
This is consistent with  previous measurements of neutral
pions~\cite{ppg028,ppg044}.  The charged kaon agrees with the charged
pion within the systematic uncertainties.  The $\phi$ meson exhibits no
apparent modification.

On the other hand, the protons show a very large and strongly centrality
dependent Cronin enhancement, reaching a factor of 2 in the most central
collisions at intermediate $p_T$.  Even in the 40--60\% centrality class
the enhancement is a factor of 1.5.  For the most peripheral bin the
enhancement is much smaller, at a factor of about 1.1--1.2, and is close
to unmodified, similar to the other particle species.  This strong
centrality dependence of the proton $R_{dA}$ is in fact very similar to
the significant centrality dependence of the $p/\pi$ ratio, and these
two observables are likely driven by the same mechanism.  Also apparent in
the $R_{dA}$ is that the enhancement for protons begins to fall off at
3.0~GeV/$c$ and steadily drops with increasing $p_T$, appearing nearly
unmodified at the highest $p_T$ points.

The $R_{dA}$ of $\pi$, $K$, $\phi$, and $p$ show significant dependence on
the number of valence quarks, and no dependence on particle mass.  That the
baryon $R_{dA}$ is quite different from that of the mesons suggests that
recombination plays a role in particle production in $d$$+$Au collisions as
well as Au$+$Au.  The kaon $R_{dA}$ is consistent with the pion $R_{dA}$, in
contrast to $R_{\rm AA}$ where the kaons are consistently above the pions.  This
is consistent with the $K/\pi$ ratio discussed above and indicates that
there is no discernible strangeness enhancement within uncertainties in $d$$+$Au collisions.

\subsection{Comparison of Peripheral Au$+$Au to Central $d$$+$Au}

Motivated by the remarkable similarities between peripheral Au$+$Au and central
$d$$+$Au collisions, we now compare the two directly.  Figure~\ref{fig:kpiratio_aud}
shows the $K/\pi$ ratio and Fig.~\ref{fig:ppiratio_aud} shows the $p/\pi$
ratio in peripheral Au$+$Au and central $d$$+$Au collisions plotted together.  In both
cases the ratios are completely consistent with each other between the different
collision species, suggesting that the particle production mechanisms in
peripheral Au$+$Au and central $d$$+$Au collisions are quite similar.

\begin{figure}[thbp]
\includegraphics[width=0.948\linewidth]{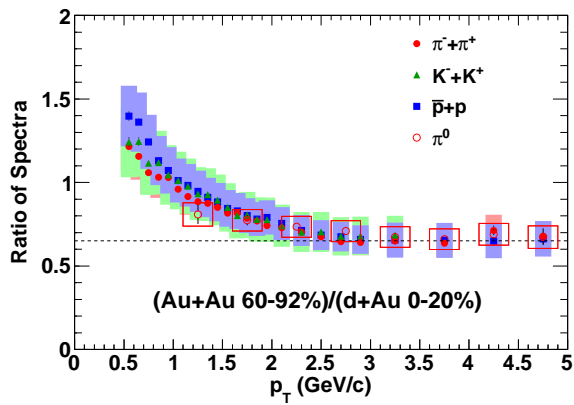}
\caption{(Color online) Ratio of invariant yield of particles
in peripheral Au$+$Au (60--92\%) to central $d$$+$Au (0--20\%)
collisions as a function of $p_T$.}\label{fig:ratio_aud}
\end{figure}

Figure~\ref{fig:ratio_aud} shows the ratio of the spectra in peripheral Au$+$Au to
central $d$$+$Au collisions for pions, kaons, and protons.  Also plotted is the ratio
for neutral pions, as determined from the data in Refs.~\cite{ppg044,ppg080}.
The ratio for $\pi^0$ shows excellent agreement with the $\pi^{\pm}$ ratios.

We note that $N_{\rm coll}$ values are within 2\% of each other and that the
$N_{\rm part}$ values are within 4\% of each other.  Both of these differences
are well within the associated systematic uncertainties of the Glauber MC
values.  No scaling is applied, but a scaling by the respective $N_{\rm coll}$
or $N_{\rm part}$ values would change the data imperceptibly.  The ratios tend
to the same value of roughly 0.65 for each particle species at and above
2.5--3~GeV/$c$.
This universal scaling is strongly suggestive of a
common particle production mechanism between peripheral Au$+$Au and central
$d$$+$Au collisions.  It is also interesting to observe that at the lower $p_T$,
where this ratio rises strongly, there is minimal mass or particle type
dependence.

Given the fact that both $N_{\rm coll}$ and $N_{\rm part}$ are essentially the same
in these two systems, any quantity or physical effect that scales with either
of these should be the same in each system, and thus should cancel almost
exactly in this ratio.  If we make the simple assumption that most or all of
the cold nuclear matter effects scale with $N_{\rm coll}$ or $N_{\rm part}$, then
those effects are completely canceled in this ratio, leaving only the hot
nuclear matter effects present in the peripheral Au$+$Au collisions.  This could
mean that this ratio being less than unity is attributable to the parton energy
loss in peripheral Au$+$Au.
This picture is consistent with the findings in this paper and elsewhere that
the $R_{\rm AA}$ of mesons indicates parton energy loss in the medium even in
peripheral Au$+$Au collisions.
It is striking, then, that this ratio is independent of particle species, which
is suggestive of similar energy loss effects even for protons.
This indicates that
the baryon enhancement mechanism is the same in both systems.

However, although $N_{\rm coll}$ and $N_{\rm part}$ are consistent for the two
systems, there is an inherent participant asymmetry that needs to be taken
into account.  In the case of peripheral Au$+$Au collisions,
one has a scenario in which 7 or 8 nucleons on the edge of one Au nucleus
collide against 7 or 8
nucleons on the edge of the other Au nucleus.  On the other hand, in the
case of central $d$$+$Au collisions,
one has a scenario in which the 2 nucleons of the
deuteron collide against 13 nucleons in the center of the Au nucleus.
This introduces several additional factors that need to be considered.  For
example, the participant asymmetry produces a rapidity shift in the particle
production~\cite{Back:2003hx}.  This may explain a deficit of soft particles
at low $p_T$ in $d$$+$Au collisions at midrapidity, which in turn would explain
why the ratio trends up at low $p_T$.
We also note this low $p_T$ region where the ratio rises is where hydrodynamics
effects are known to be important in Au$+$Au collisions.
It is possible that there are collective flow effects in $d$$+$Au collisions as well,
as suggested by the recent results reported in
Refs.~\cite{ppg152,CMS:2012qk,Abelev:2012ola,Aad:2012gla}.
A full viscous hydrodynamics model comparison is warranted.

Another issue to consider is the modification of parton distribution
functions (PDFs) in nuclei.  These nuclear PDFs (nPDFs) are known to be
modified from the PDFs of single nucleons~\cite{Arneodo:1992wf,Eskola:2009uj}.
The
experimentally measured nPDFs are averaged over the entire nucleus and are
typically compared to the deuteron PDF to determine the nuclear modification.
The binary collisions in peripheral Au$+$Au involve two nucleons which have
modified nPDFs.  On the other hand, the binary collisions in central $d$$+$Au
involve an approximately unmodified nucleon from the deuteron and a modified
nucleon from the Au nucleus.
Physical observables sensitive to the nPDFs
would then be expected to be different for the two systems.
However, it is possible that the nucleons in the more diffuse outer region
of the nucleus have a different modification from those in the denser center.
Therefore, it is not possible to make any model-independent quantitative
statements about the
differences between the nPDFs in these two systems.

\section{Summary}
\label{s:summary}

In summary we present a highly detailed and systematic study of identified charged
hadron spectra and ratios as a function of $p_T$ and centrality for Au$+$Au and $d$$+$Au
collisions at $\sqrt{s_{_{NN}}}$~200~GeV.
As has been reported previously, we find a baryon enhancement present in both
systems.  In $d$$+$Au collisions, the Cronin enhancement has long
been known to be stronger for baryons than for mesons.  However, for the first
time a study with enough statistical and systematic precision presents clear
evidence for a strong centrality dependence of this effect.
In Au$+$Au collisions the baryon enhancement has been attributed to parton
recombination as the mode of hadronization.  A version of the recombination model
has been applied to $d$$+$Au collisions as well, which reproduces the baryon vs.
meson differences.  The present data strongly suggest that further theoretical 
investigation is warranted.
Given the excellent statistical precision of the present data set, a direct
comparison between the two is made for the first time.  Specifically, a ratio of
the spectra in the most peripheral Au$+$Au and most central $d$$+$Au collisions is
measured.  These two systems have nearly identical values of both $N_{\rm coll}$ and
$N_{\rm part}$.  Therefore, a direct comparison between the two cancels out a
large number of physical effects. 
We conclude that the baryon
enhancement present in both systems is likely driven by a common hadronization
mechanism.


\begin{acknowledgments}

We thank the staff of the Collider-Accelerator and Physics
Departments at Brookhaven National Laboratory and the staff of
the other PHENIX participating institutions for their vital
contributions.  We acknowledge support from the 
Office of Nuclear Physics in the
Office of Science of the Department of Energy, the
National Science Foundation, Abilene Christian University
Research Council, Research Foundation of SUNY, and Dean of the
College of Arts and Sciences, Vanderbilt University (U.S.A),
Ministry of Education, Culture, Sports, Science, and Technology
and the Japan Society for the Promotion of Science (Japan),
Conselho Nacional de Desenvolvimento Cient\'{\i}fico e
Tecnol{\'o}gico and Funda\c c{\~a}o de Amparo {\`a} Pesquisa do
Estado de S{\~a}o Paulo (Brazil),
Natural Science Foundation of China (P.~R.~China),
Ministry of Education, Youth and Sports (Czech Republic),
Centre National de la Recherche Scientifique, Commissariat
{\`a} l'{\'E}nergie Atomique, and Institut National de Physique
Nucl{\'e}aire et de Physique des Particules (France),
Bundesministerium f\"ur Bildung und Forschung, Deutscher
Akademischer Austausch Dienst, and Alexander von Humboldt Stiftung (Germany),
Hungarian National Science Fund, OTKA (Hungary), 
Department of Atomic Energy and Department of Science and Technology (India), 
Israel Science Foundation (Israel), 
National Research Foundation and WCU program of the 
Ministry Education Science and Technology (Korea),
Ministry of Education and Science, Russian Academy of Sciences,
Federal Agency of Atomic Energy (Russia),
VR and Wallenberg Foundation (Sweden), 
the U.S. Civilian Research and Development Foundation for the
Independent States of the Former Soviet Union, 
the US-Hungarian Fulbright Foundation for Educational Exchange,
and the US-Israel Binational Science Foundation.

\end{acknowledgments}



\end{document}